\documentclass[12pt,a4paper]{article}
\pdfoutput=1
\usepackage{amsmath}
\usepackage{amsthm,amssymb}
\usepackage{rotating}
\usepackage{pdflscape}
\pdfoutput=1 
\numberwithin{equation}{section}
\usepackage{graphicx}

\title{\Large \bf    Correlations between submission and acceptance of papers in  peer review  journals 
 }

\author{  \large \bf  Marcel    Ausloos$^{1,2}$, Olgica Nedic$^{3}$, Aleksandar Dekanski$^{4}$
  \\ \\$^{1}$  School of Business,  College of Social Sciences, Arts, and Humanities, \\University of  Leicester, Leicester,   LE1 7RH, United Kingdom \\  email:  ma683@le.ac.uk
 \\$^{2}$   Group of Researchers for Applications of Physics \\in Economy and Sociology (GRAPES), \\Rue de la belle jardini\`ere, 483, B-4031 Angleur, Li\`ege, Belgium \\  
 email: marcel.ausloos@ulg.ac.be 
 \\$^{3}$  Institute for the Application of Nuclear Energy (INEP),\\
University of Belgrade,  Banatska 31b, Belgrade-Zemun, Serbia 
  \\$^{4}$  Institute of Chemistry, Technology and Metallurgy, 
\\Department of Electrochemistry,\\ University of Belgrade,   Njegoseva12,  Belgrade, Serbia  }

\begin{document}
\maketitle  
\newpage
\begin{abstract}  
This paper  provides a  comparative study  about seasonal  influence on editorial decisions for papers submitted to   peer review journals. We distinguish  a specialized one, the Journal of the Serbian  Chemical  Society ({\it JSCS}) and an interdisciplinary one, {\it Entropy}. 
 Dates of  electronic submission for  about 600 papers to  {\it JSCS}  and 2500 to {\it Entropy} have been recorded    over  3 recent years.
 Time series  of  either accepted or rejected papers are subsequently analyzed.   We take either editors or authors view points into account, thereby considering  magnitudes and probabilities.
 
It is found that there are distinguishable peaks and dips in the time series, demonstrating  preferred months   for  the submission of papers.  It is also found that papers are more likely accepted   if they are submitted during  a few specific months, -  these depending on the journal.
 The probability  of having a rejected paper  also appears to be seasonally biased. 
 
In view of clarifying  reports with contradictory findings, we discuss previously proposed conjectures for such effects, like holiday effects and  the  desk rejection by editors. We  conclude that  the type of journal, specialized or multidisciplinary, seems to be the drastic criterion for distinguishing the outcomes rates.

\end{abstract}

\newpage
\section{Introduction}\label{introduction}

 In the peer review process,  two "strategic" questions have to be considered: on one hand, - for editors, what is the load due to the number (and  what is the relative frequency) of papers submitted at some time during a year?;  on the other hand, - for authors, is there any bias in the probability of acceptance of their (by assumption  high quality) paper  when submitted in a given month, because of the (being polite) mood of editors and/or reviewers?
   A study about such a time concentration (and dispersion) of submitted papers and their subsequent acceptance (or rejection)  seems to become appropriate  from a scientometrics point of view,  in line with recent "effects" found and known through media, like coercive citations or faked research reports.
 
In fact, the mentioned question of paper submissions  timing is  of  renewed interest nowadays  in informetrics and bibliometrics due to the flurry of new publication journals  by  electronic means.  Moreover, 
paper acceptance  rate is of great concern to authors who feel much bias at some time. No need to say that the peer review process is sometimes slow, with reasons found in editor's and reviewers' workload,  whence a difficulty of finding reviewers. 
 Tying such questions  are  the open access policy and the submission fees imposed by publishers. on one hand, but also doubts or constraints about the efficiency in managing peer-review of scientific manuscripts–editors’ perspective  \cite{Nedicefficiency} and of authors \cite{Nedicauthorsperspective}. Thus, one may wonder if there is some "{\it seasonal"} or {\it "day of the week"} effect.

 Very recently,  Boja et al. \cite{SCIMHerteliu}, in this journal,    showed that "the {\it day of the week} when a paper is submitted to a peer reviewed journal correlates with whether that paper is accepted", when looking at a huge set of cases for high Impact Factor journals. However, there was no study of rejected papers.
 
 From the {\it seasonal} point view, previously, but in   recent time, Shalvi et al.  \cite{r11}  discussed the case of electronic submission  monthly frequency  to two psychology journals, Psychological Science  (PS) and  Personality and Social Psychology Bulletin (PSPB),  over 4 and 3 years respectively.  Shalvi et al.  \cite{r11}  found a discrepancy in the pattern of "submission-per-month" and "acceptance-per-month" for PS, - but not for PSPB.  
More papers were submitted to PS during "summer months", but no  seasonal bias effect  (based on  a $\chi^2_{(11)}$ test for percentages)   was found about subsequent acceptance; nevertheless, the percentage of accepted papers when submitted in Nov. and Dec.  was found to be very low.  In contrast, many papers were submitted  to PSPB during "winter months",   followed by a dip in April, but the percentage of  published  papers was found to be  greater if the submission to  PSPB occurred in  [Aug.-Sept.-Oct.]. Moreover,  a  marked  "acceptance success dip" occurred if the submission was in "winter months".
The main difference  between such patterns was   
conjectured to stem  from different rejections policies 
 i.e. employing desk rejections or not.  
 
Later, Schreiber   \cite{r10} examined   
  submissions to  a specialized journal, Europhysics Letters (EPL),  over 12 years.  
  He observed that  the 
  number of submitted manuscripts had been steadily increasing while  the number of accepted manuscripts had grown more slowly.
He  claimed to find   no  statistical effect. However,  - from Table 2 in  \cite{r10}, there is  a clearly visible  maximum for the number of submissions  in July,   more than 10\% over the yearly mean, and a marked dip  in submissions in February, - even  taking into account the "month  small length".  Examining the acceptance rate (roughly ranging between  45 and 55 \%, according to the month of  submission), he  concluded  that  strong fluctuations can be seen,  
  between different months,.  
  One detects a maximum  in  July and a minimum in January for the most recent years.

Alikhan  et al.  \cite{r1}  had a similar concern:  
  they  compiled   
  submissions, in 2008,   to 20  journals  pertaining to dermatology. It was  found that May was the
 least popular month,  
  while July was the most popular month.  
  We have estimated  a $\chi^2 \simeq  36.27$,  from  the  Fig. 1 data in   Alikhan et al.  \cite{r1}.
 thereby suggesting a far from uniform distribution.  
 There is no information on acceptance rate in  \cite{r1}.

  Other papers have appeared  pretending  discussing   seasonal or so effects, concluding from fluctuations, but finding  no effect,  from  standard deviations arguments, -  instead of $\chi^2$ tests. Yet, it  should be obvious to the reader that a $\chi^2$ test performs better in order to find whether a distribution is uniform or not, -  our research question. In contrast, a test based on the standard deviation and the confidence interval can only allow some claim  on some  percentage deviation of (month) outliers; furthermore  such studies are  tacitly assuming a normality of the (submission or acceptance) fluctuation distributions, - which is far from being the case.  Usually, the skewness and kurtosis of the distributions to be mandatory complements are not provided in  "fluctuations studies" by such authors. 
  
In order to contribute answers  to the  question on "monthly  bias", we have been fortunate to get access to   data  for submitted,  and later either accepted or rejected, papers to a specialized (chemistry) scientific journal and to a multidisciplinary journal. Two coauthors of the present report,  ON and AD, are    Sub-Editor  and  Manager of the   Journal of the Serbian Chemical Society  ({\it JSCS}).  One coauthor  MA is a member of the editorial board of  {\it Entropy}. It will be seen that comparing features from these two journals allows one  to lift some veil on the reported apparent discrepancy in other cases.

  Thus, here below, we explore the fate  of  papers    submitted for peer review during a given month, plus their publication fate.  We find that, in the case at hands,   fluctuations of course  occur from one year to another. However,   for $JSCS$, submission peaks  do occur in July and September,   while many less papers are submitted in May and December. A marked dip in submissions occurs in August  for $Entropy$, - the largest number of submissions occurs  in October and December.
  
 However, if the number of submitted papers is  relevant for editors and handling machines, the probability of acceptance (and rejection)  is much concerning authors. Relatively  to the  number of submitted papers, 
 it is  shown that more papers are accepted for publication  if they are submitted in January (and February), - but less so if submitted in December,  for  {\it JSCS}; the highest rejection rates  occur  for papers submitted in  December  and March.    For $Entropy$,  the  acceptance rate is the lowest in June and December, but is high for papers submitted during spring months, February to May.  
 Statistical tests, e.g., $\chi^2$ and confidence intervals,   are provided to ensure the validity of the findings.   
  
  Due to different desk rejection policies and in order to discuss the effect of such policies as in \cite{r1}, we discuss a possible specific determinant for {\it JSCS} data: possible effects due to religious or holiday   bias  (in Serbia) are commented upon.

\section{Data}\label{Data}

The  {\it JSCS}  and {\it Entropy} peer review process are both mainly managed electronically, - whence the editorial work is  only weakly  tied to the  editors working days\footnote{N.B.  Nevertheless,  there are days of the week  effects    \cite{r2}.  }. 
 
\subsection{The  Journal of the Serbian Chemical Society  }\label{dataJSCS}

{\it JSCS}  contains 14 sub-sections and many sub-editors, as it   can be viewed from the journal website
 $http://shd.org.rs/JSCS/$.

\begin{figure} 
 \includegraphics[height=9.8cm,width=7.8cm]
{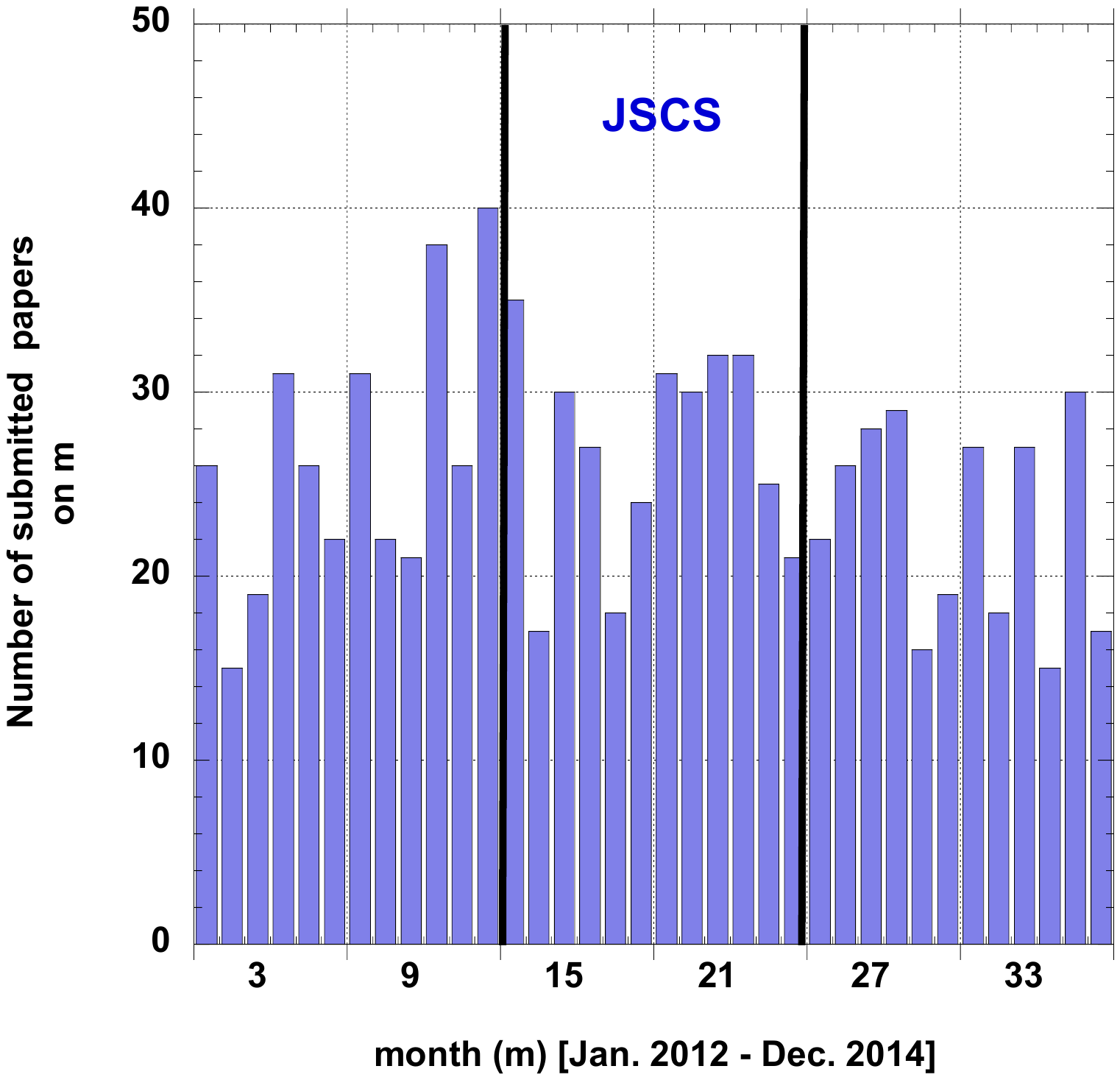}  
  \includegraphics [height=9.8cm,width=7.8cm]
{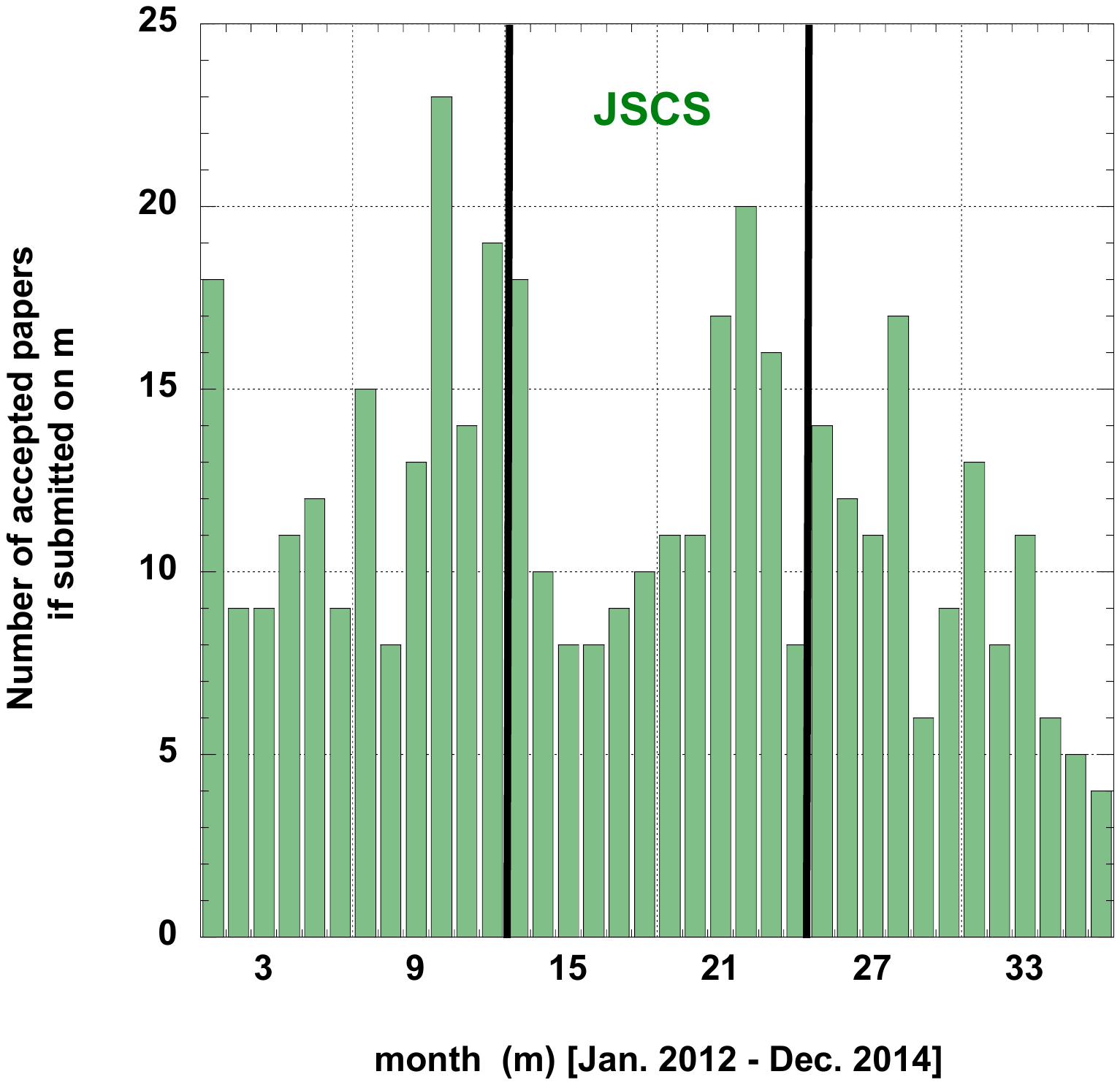}  
 \caption[ ] { Time series   of (left) the number  of \underline{submitted}   papers
   and (right) of the number of \underline{accepted  papers when submitted} to   {\it JSCS} during a given month ($m$)  in   2012,  2013  and  2014. }
  \label{Plot24NsNatimeseries} 
 \end{figure}

 The   (36 data points) time series of the monthly submissions $N_s^{(m,y)}$  to  {\it JSCS} in a given month ($m= 1,\dots,12$) in   year ($y$) [2012, 2013 and 2014]  is shown in Fig. \ref{Plot24NsNatimeseries}.  
 The total number of submission ($T_s^{(y)}=\sum_m  N_s^{(m,y)}$) decreased  by $\sim 17\%$ or so from $y$=2012 or 2013 to $y$=2014: 317 or 322$\rightarrow 274$.  
    
Next, let us call  the numbers of papers later accepted ($N_a^{(m,y)}$) and those rejected ($N_r^{(m,y)}$).    
 Among the  total number of submitted papers ($T_s=\sum_y T_s^{(y)} $)= 913 submitted papers,  $T_a= 424 $ (= 162 + 146 + 116)   were finally accepted for publication. 
   In view of further discussion, let it be pointed out that among the total number  $T_r$ = 474  (=  149 + 172 +153)  of  ({\it peer} and subsequently $editor$) rejected papers, i.e., 52\%,  $T_{dr} =    ( 42 + 81 + 79=) $  202 papers   were {\it desk rejected},  without going  to a peer review process, i.e. 22.1\%. 
 For completeness, let it be recorded that   several papers     were rejected because  the authors did not  reply to the reviewers remarks in due time and  a few submissions were withdrawn.  
    (Thus, $T_a+ T_r \neq T_s$:   424 + 474   $\neq$ 913).
 
 The time series of the positive  fate, thus acceptance,  of submitted  papers  for  a specific month  submission is also shown in Fig.   \ref{Plot24NsNatimeseries}. 
 
 The statistical characteristics\footnote{Thereafter, the indices $m$ and $y$ are not written, for simplicity, if there is no ambiguity.} of the $N_s^{(m,y)}$, $N_a^{(m,y)}$, $N_r^{(m,y)}$, and  $N_{dr}^{(m,y)}$   distributions for $JSCS$ are given in  Table  \ref{JSCSNsmonthstat} -   Table \ref{JSCSNdrmonthstat}.

 \subsection{Entropy  }\label{dataEntropy} 

 {\it Entropy}  covers research on all aspects of entropy and information studies.   The journal home page is
 $http://www.mdpi.com/journal/entropy$. 
 
\begin{figure} 
 \includegraphics[height=6.8cm,width=6.8cm]
{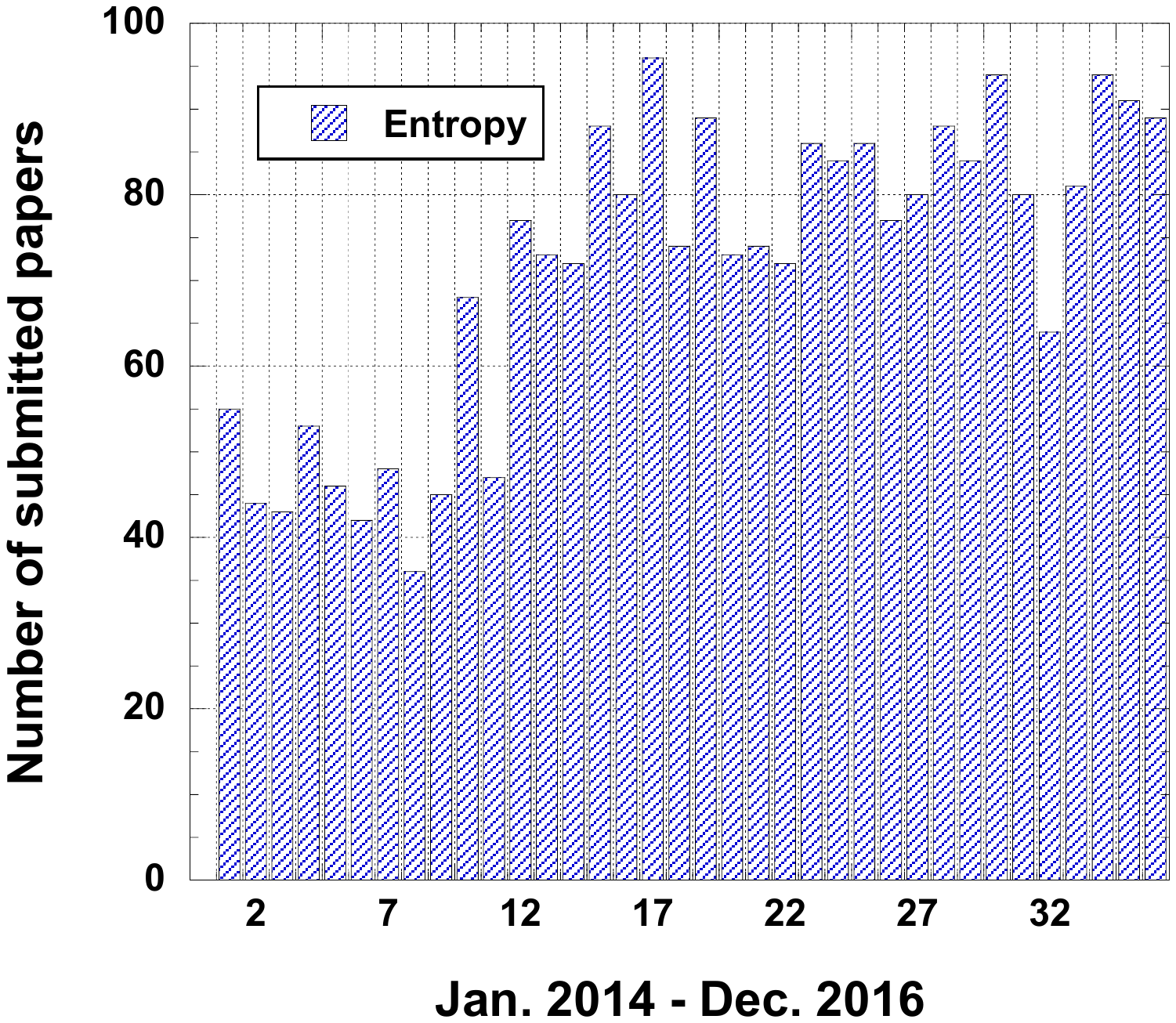}  
  \includegraphics [height=6.8cm,width=6.8cm]
{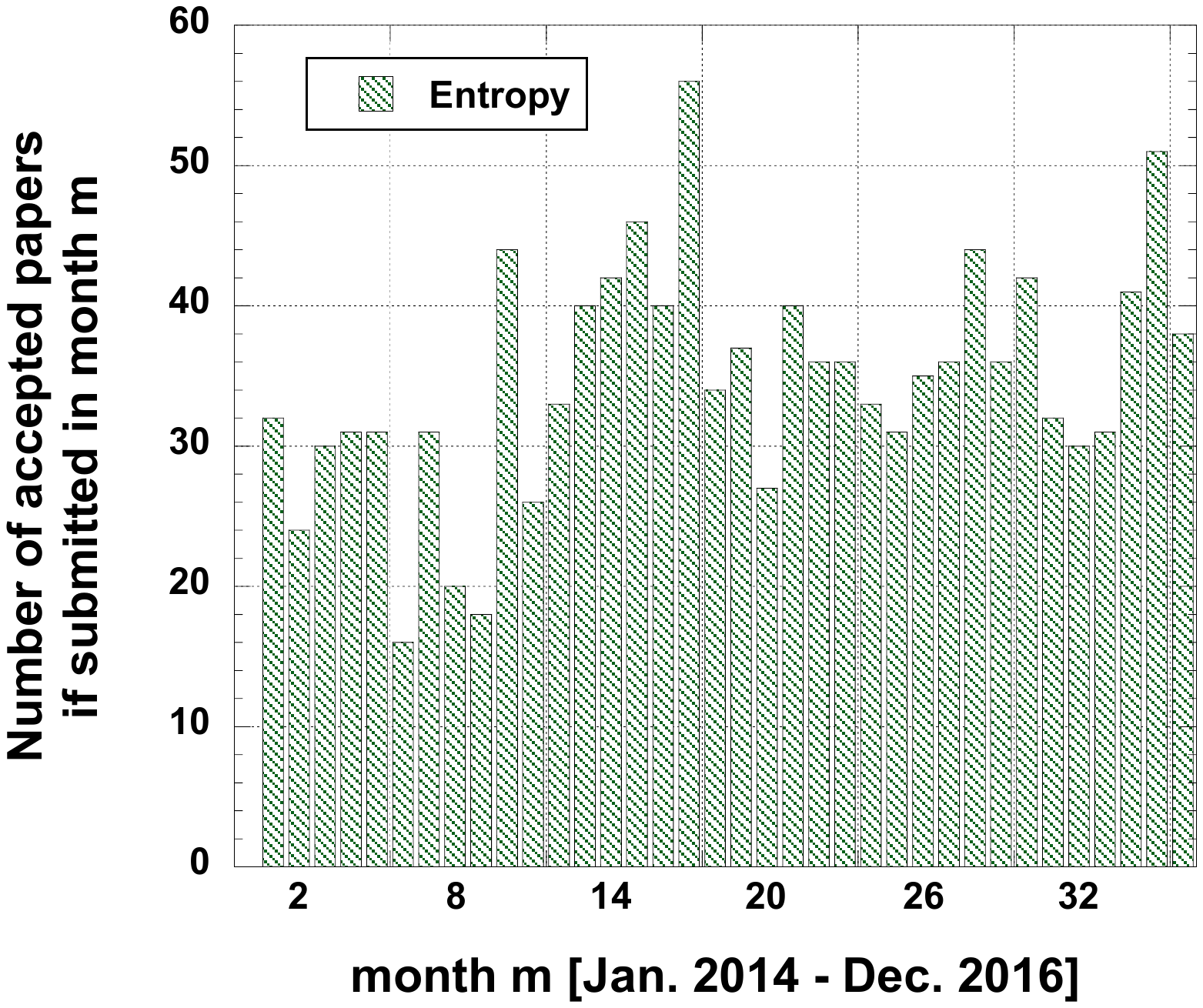} 
 \caption[ ] { Time series of (left) the number  of \underline{submitted}   papers
   and (right) of the number of \underline{accepted  papers when submitted} to   {\it Entropy}    during  a given month in 2014, 2015,  and 2016}
  \label{Plot8EntropyNsNatimeseries} 
 \end{figure}

 The   (36 data point) time series of the monthly submission  to  {\it Entropy} over the years 2014, 2015,  and 2016 is shown in Fig. \ref{Plot8EntropyNsNatimeseries}.    The number of submission increased  by $\sim 60\%$ or so from 2014 to 2015: 604$\rightarrow 961$, but not much ($\sim 5\%$)  between 2015 and 2016:   
961$\rightarrow1008$.   

 Among the  $T_s=2573$ submitted papers,  $T_a= 1250 $ were finally accepted for publication. 
  The time series of the positive  fate, thus acceptance,  of submitted  papers  after  a specific month  submission, is also shown in Fig.  \ref{Plot8EntropyNsNatimeseries}.  
 
  In view of further discussion below, let it be pointed out that   there were  (110 + 162 + 246	=)  518 peer review rejected papers,  i.e.  20.1\%;  $T_{dr} =  
 (158 + 332 + 315 =) $ 805   papers were desk rejected at submission, i.e. 31.2\%.    
 
 The statistical characteristics of the  $N_{s}^{(m,y)}$, $N_{a}^{(m,y)}$,   $N_{r}^{(m,y)}$,   and  $N_{dr}^{(m,y)}$  distributions  for $Entropy$ are given in  Table  \ref{EntropyNsmonthstat} -   
 Table \ref{EntropyNdrmonthstat}.
 
 \section{Data analysis}\label{analysis}

   The most important value to discuss is  the calculated $\chi^2$, for checking whether or not the distribution is uniform over the whole year.
 
 Notice that we can discuss the data not only comparing different years, but also the cumulated data: $C_s^{(m)}=\sum_y N_s^{(m,y)}$, and similarly for  $C_a^{(m)}$, $C_r^{(m)}$, and $C_{dr}^{(m)}$, as if all years are "equivalent".
  For further analysis,  we provide the statistical characteristics of the cumulated distributions in  
   Table  \ref{JSCSNsmonthstat}  - Table \ref{EntropyNdrmonthstat}.
   
 We have also taken into account   that months can have a different number of days, normalizing all months as if there were 31 days long (including the special case of February  in 2016). The fact that the number of papers appears not to be an integer, in so doing,  is not a drastic point, but more importantly such a data manipulation does not disagree at all with our following conclusions.  Thus, we do not report    results  due to such "data   normalization". 
 
  \subsection{{\it JSCS} Data analysis}\label{analysisJSCS}
  
 In all {\it JSCS} cases, the mean of each distribution decreases from 2013 to 2014;   so does the standard deviation $\sigma$. This is the case for  the cumulated time series, $C^{(m)} $ = $N^{(m,2012)} +  N^{(m,2013)}  + N^{(m,2014)}  $, data which necessarily differs from   $N^{(m,[2012-2014])}$.  The coefficient of variation (CV $ \equiv \sigma/\mu$)   is always quite small, indicating  that the data is reliable beyond statistical  sampling errors.  Each     coefficient of variation\footnote{The coefficient of variation  is usually used to compare distributions; even if the means are drastically different from one another; its value also points toward a possible anomalous spread of the distribution or a multipeak effect. }, $C_s$ or $C_a$ or  $C_r$ or $C_{dr}$,  for the  cumulated data is lower than  the other CVs; this is  {\it a posteriori} pointing to the  (statistical) interest of accumulating  data for each month of different years, - beside looking at the more dispersed data over a long time span.

 Next,  observe  the summary of statistical characteristics  in  Table \ref{JSCSNsmonthstat} - 
Table \ref  {JSCSNdrmonthstat}; they    show that the distributions are positively skewed,  except   those for the submitted  papers  
which are negatively skewed.   The kurtosis of each distribution is  usually  negative,  except for the anomalous cases,  $N_r^{(m,2014)}$ and  $N_{dr}^{(m,2014)}$, whence for the latest case for the whole series.  It can be concluded that the distributions are  quite asymmetric, far from a Gaussian, but rather peaked.

 Almost all measured values fall within  the classical confidence interval  $ ] \mu-2\sigma,\mu+2\sigma[$. However,  in five cases,   a few extreme values fall above the upper limit, as can be deduced from the Tables.    
 
"Finally", notice that all $\chi^2$  values, reported in Table  \ref{JSCSNsmonthstat}  
- Table  \ref{JSCSNdrmonthstat} are much larger than the 95\% critical value:  they markedly allow to reject the null hypothesis, i.e. a uniform distribution, for each examined case.  Thus a monthly   effect exists beyond statistical errors \underline{for all} $N_s$, $N_a$,   $N_r$  and $N_{dr}$ cases.  
 
  \subsection{{\it Entropy} Data analysis}\label{analysisENTROPY}
  
  In the case of {\it Entropy} data,  the CV is usually low;, - and much lower than in the case of {\it JSCS}. The skewness and kurtosis  are not systematically positive or negative. The number of outliers outside the confidence interval is also "not negligible"; this is  hinted  from the number of maximum  and minimum values falling outside the confidence interval, yet "not too far" from the relevant interval border. Nevertheless, this implies that the distribution behaviors are influenced by the number of data points,  to a larger extent  for {\it Entropy} than for {\it JSCS}.
  
  Nevertheless,  notice that all $\chi^2$  values, reported in Table  \ref{EntropyNsmonthstat}  
- Table  \ref{EntropyNdrmonthstat} are also much larger than the 95\% critical value:  they markedly allow to reject the null hypothesis, i.e. a uniform distribution, for each examined case.  A  month anomaly effect exists beyond statistical errors \underline{for all} $N_s$ and  $N_a$; it is weaker for  the $N_r$  and $N_{dr}$ cases.  The large  $\chi^2$ values obviously point to  distinguishable peaks and dips,  thereby markedly promoting the view of monthly effect bias  for $N_s$ and  $N_a$. 
  
\section{Discussion}\label{reasoning}

 Let us first recall that the journals here examined have different aims; one is a specialized journal, the other is an interdisciplinary journal. To our knowledge, this is the first time that  a journal with such a "broadness"  is considered within the question on monthly bias. It seems that one should expect an averaging effect due to a varied number of constraints on the research schedules pertaining to different topics and data sources. One subquestion pertains on whether a  focussed research influences the timing of paper submission, and later acceptance (or rejection). One would expect more bias for the {\it JSCS} case than for the {\it Entropy} case.
Comparing    journals (in psychology), but with different "specializations", Shalvi et al. \cite{r11} had found different behaviors indeed.  
   Let us observe what anomalies are  found  in the present cases. 
   
\subsection{JSCS}\label{JSCSreasoning}
 
Comparing months in 2012, 2013 and 2014, it can be noticed that the most similar months (the least change of positions in the  decreasing  order of "importance") are Dec., May, June  for the lowest submission rate,  while Sept. and July are those remaining on the top of the month list, for the highest submission rate; see figures.    A specific deduction  
seems to be   implied:  there is a steady academic production of papers strictly before and after holidays, but there is a quiet (production and) submission  of papers before holidays.  This finding  of  July production  relevance is  rather similar to that found for  most other journals, - except  PSPB 
   \cite{r11}.  
   
   
    Concerning the May dip anomaly, one may remind ourselves that  in most countries (including Serbia),
lectures and practical work at faculties end by June; since
many authors (professors, assistants) are very engaged with
students at that time,  probably May is not the month when they
are focused on writing papers but rather "prefer" finishing regular duties. In fact,  corroborating this remark, it has been observed that most papers submitted to {\it JSCS} are from academia researchers 
 \cite{r9}.
 
A huge peak in January 2013 is intriguing. It was  searched whether something special  occurred $ca.$ January 2013;  it was checked  that the submission system worked properly:  there was no special clogging a month before.  Moreover,  there were no special invitations or collection of articles for a special issue.  Therefore, the peak can be correlated to that found for PS. From such a concordance, it seems that more quantitative correlation aspects could be searched for through  available data. 
   
    Notice that on a  month rank basis,  for 2013 and 2014, the Kendall $\tau$ coefficient  $\simeq -0.0303$  for submitted papers, but  $\simeq -0.3030$ for accepted papers; concerning the correlation between the cumulated $N_s$ and $N_a$,  the Kendall $\tau$ coefficient $\simeq -0.2121$. 
    
   Two other points related    to {\it JSCS},  are discussed  in Sect. \ref{deskrejection} and \ref{authorschance}: (i)  the  possible influence of desk rejection policy, a conjecture of Shalvi et al.   \cite{r11},  for distinguishing patterns, and (ii) the acceptance and rejections rates, which are tied to the submission patterns, but also pertain  to  the "entrance barrier"  (editor load mood) conjecture  proposed by Schreiber   \cite{r10}.

\subsection{Entropy}\label{Entropyreasoning}
    
In the case of {\it Entropy}, the cumulated measure (over the 3 years here examined)  points to a more frequent submission in December, and a big dip in August. From a more general view point, there are more papers submitted  during the last  3 months of the year. A marked contrast occurs for  the accepted papers  for which  a wide dip exists  for 4 months : from June till September.  The discussion on  desk rejection and better chance for acceptance are also found  in Sect. \ref{deskrejection} and \ref{authorschance}.

Notice that  for  the correlation between the cumulated $N_s$ and $N_a$,   the Kendall $\tau$ coefficient $\simeq  0.4242$.

Finally, comparing the cumulated numbers of submitted and accepted papers to {\it JSCS} and to {\it Entropy}, and ranking the months accordingly, the Kendall $\tau$ coefficient is weakly negative: $\simeq$ -0.333 and -0.1818, respectively.

\section{Constraint determinants}\label{constraints}
\subsection{Seasonal  desk rejection  by editor   
    }\label{deskrejection}

Often controversial or scorned upon, the desk rejection patterns at {\it JSCS} and {\it Entropy} can be discussed now.   Table  \ref{JSCSNdrmonthstat}  and Table \ref{EntropyNdrmonthstat} provide  the relevant data   respectively.   
Notice that  for either  $JSCS$ or  $Entropy$, we do not discuss reasons  why  editors (and reviewers)  reject  papers; these reasons  are outside the present considerations; see  for some information   \cite{r3,r4,r6}. 
    
\begin{figure} 
\includegraphics [height=7.8cm,width=7.8cm]
{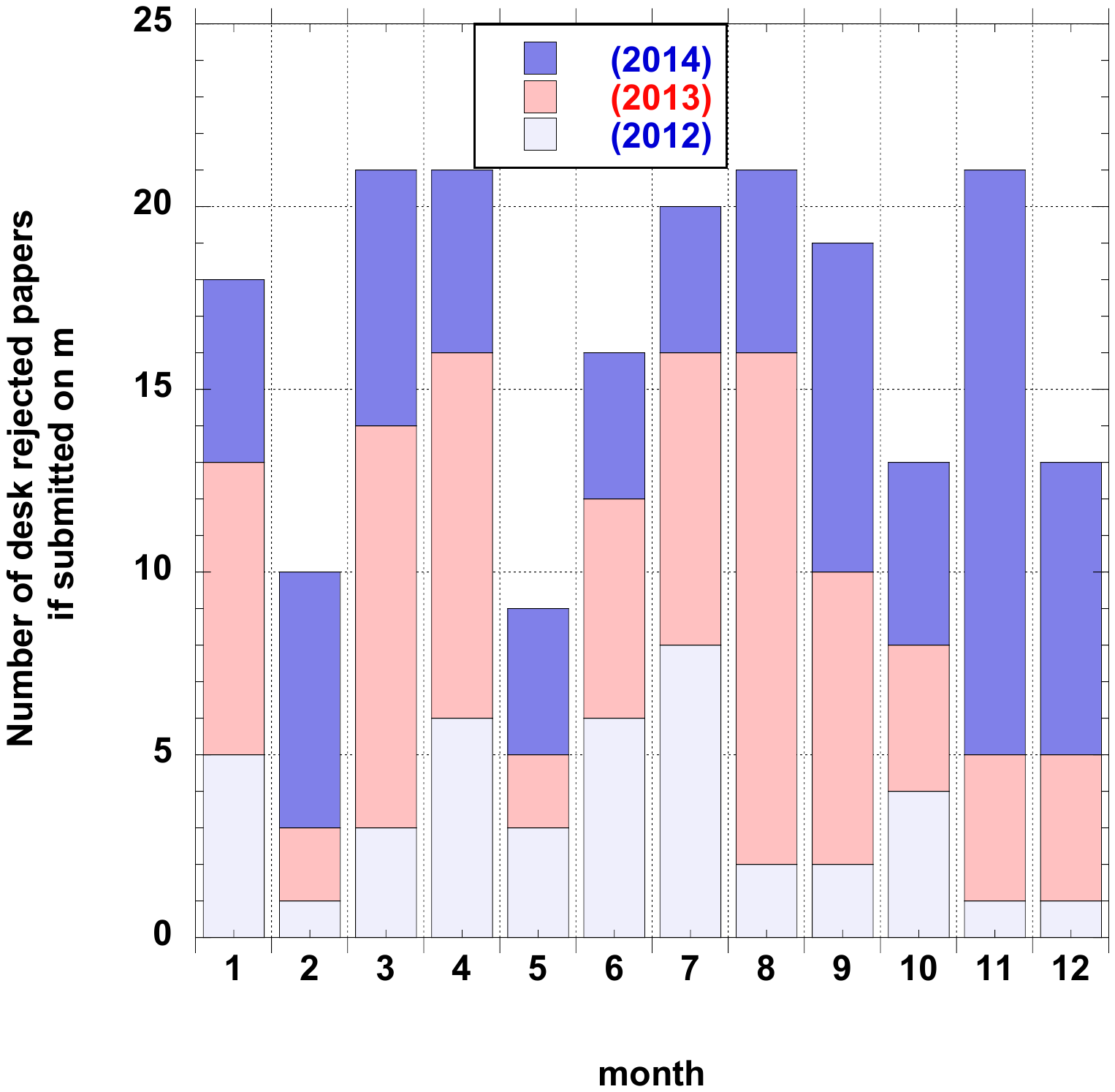}  
     \includegraphics [height=8.8cm,width=7.8cm]
{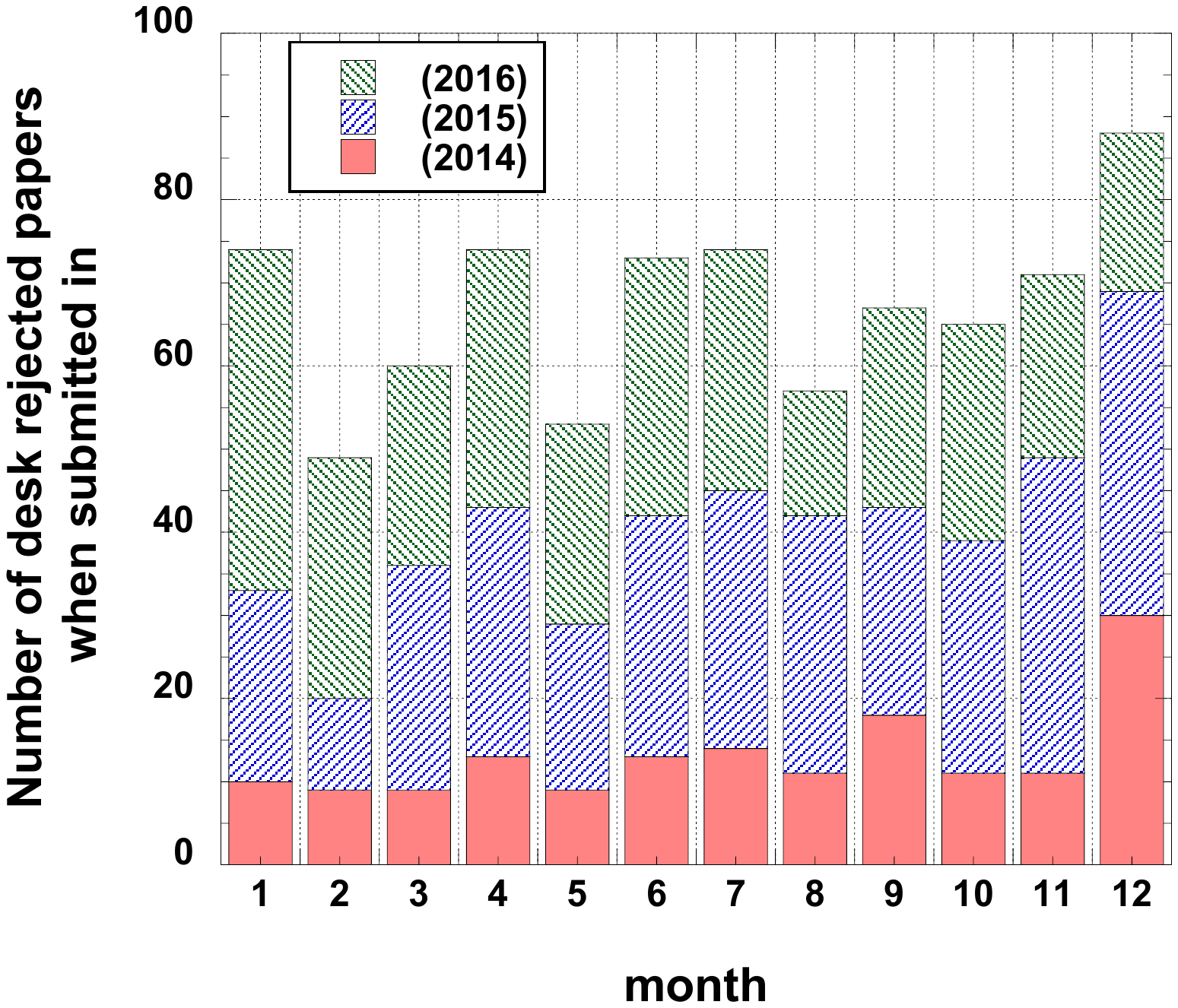}  
 \caption{  Aggregated distribution of the number of  \underline{desk rejected} papers when submitted 
 (left) to  {\it JSCS} during a given month, in 2012  or in 2013 or in 2014 and  (right) to  {\it Entropy}  during a given month in  2014 or  in 2015,  or  in 2016. } 
 \label{Plot10Ndr13Ndr14} 
\end{figure}

    Let us consider $JSCS$ first. It can be observed that  "only"  (160/596) $\simeq$  27\% papers are desk rejected, - this is interestingly compared to the ($"many"$) rejected papers after peer review: 325/596 $\simeq 0.55$, for {\it JSCS}; the ratio is  $\sim 1/2$. The highest desk rejection rate occurs for papers submitted in Nov., while the lowest is for those submitted in  May; see Fig. \ref{Plot10Ndr13Ndr14}.  Distinguishing  years, it happens that a high rejection rate occurs if the papers were submitted in Nov. 2014 and Aug.  2013, while a low rejection rate occurred for papers submitted in Feruary  and May 2013. 
   
There is no apparent month correlations.   For example, the month with the greatest
number of submissions (overall) is Sept.; the rejection rate in Sept. 2013 was 0.469, out of which 0.250 were
desktop rejected. In Sept. 2014, these percentages were 0.555 and 0.333. On the other hand, the month with the lowest
number of submissions is May. In May 2013, the rejection rate was 0.500, but desktop rejection was only  0.111. In May 2014, the  rejection rate was
0.562, and desktop rejection was  0.250.  
     
     For completeness in arguing, let it be known that   official holidays in Serbia are on  
Jan. 1-2 and 7 (Christmas day), Feb.  15-16, in April (usually) one Friday and one Monday (Easter holiday), May
1-2,  and Nov. 11, - at which time one does not expect editors to be on duty for desk rejection.

Next concerning {\it Entropy},  (805/2573) $\simeq$  31\% are desk rejected at submission, much more than those rejected by reviewers (and the editor), i.e.
(518/2573) $\simeq$ 20\%. The greatest desk rejection occurs in December and January, - the lowest in February, May, and August. However, in terms of percentage of desk rejection  with respect to the number of submitted papers, the months of December, September and June are the most probable, while in February and May  the editors seem more soft.

Conclusions: 
 there seems  to be no effect due to holidays on the editorial workflow, as months most often  containing holidays (January, July
and August) exhibit no special statistical anomaly, -   with respect to either submission or decision rate as compared to other
months, for {\it JSCS}. Yet, the $\chi^2$ is quite  large ($\sim$16.55; see Table  \ref{JSCSNdrmonthstat}). Thus, the seasonal effect might have another origin.   The {\it Entropy} $N_{dr}$ data  distribution is even more uniform
($\chi^2 \sim$ 6.52; see Table \ref{EntropyNdrmonthstat}).   
If any, some seasonal effect on  $N_{dr}$  might occur during winter time.

\subsection{Entrance barrier editor load effect}

Schreiber \cite{r10} considers that an entrance barrier can be set up by editors due to their work load. We understand such a bias as resulting from an accumulation of submitted papers at some time thereafter correlated to a large rate of desk rejection. One can without much restriction assume that the correlation has to be observed for zero month-time  lag, since both journals are usually prone to replying quickly to authors.   

\begin{figure} 
\includegraphics [height=7.8cm,width=7.5cm]
{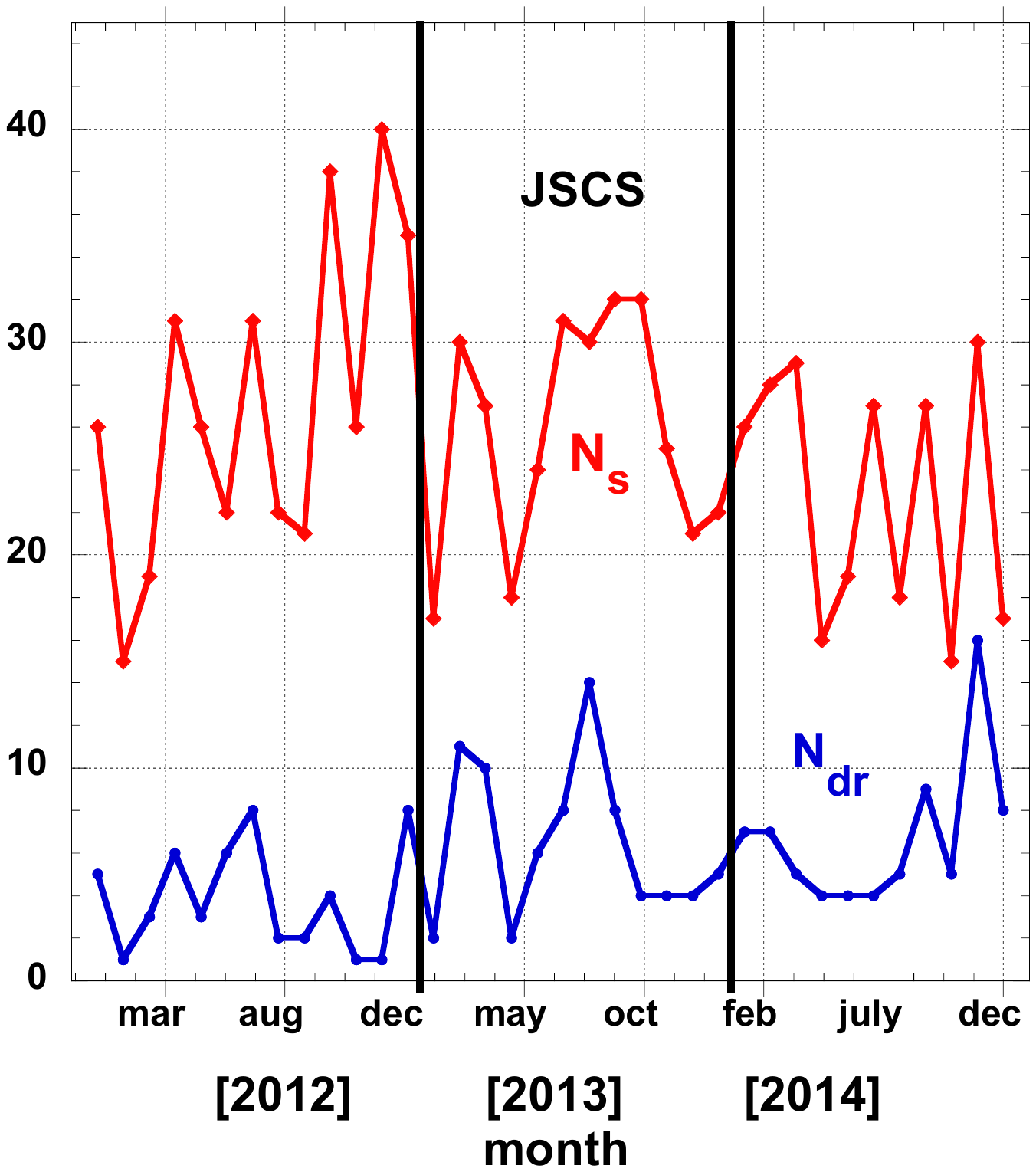}  
     \includegraphics [height=7.8cm,width=7.50cm]
{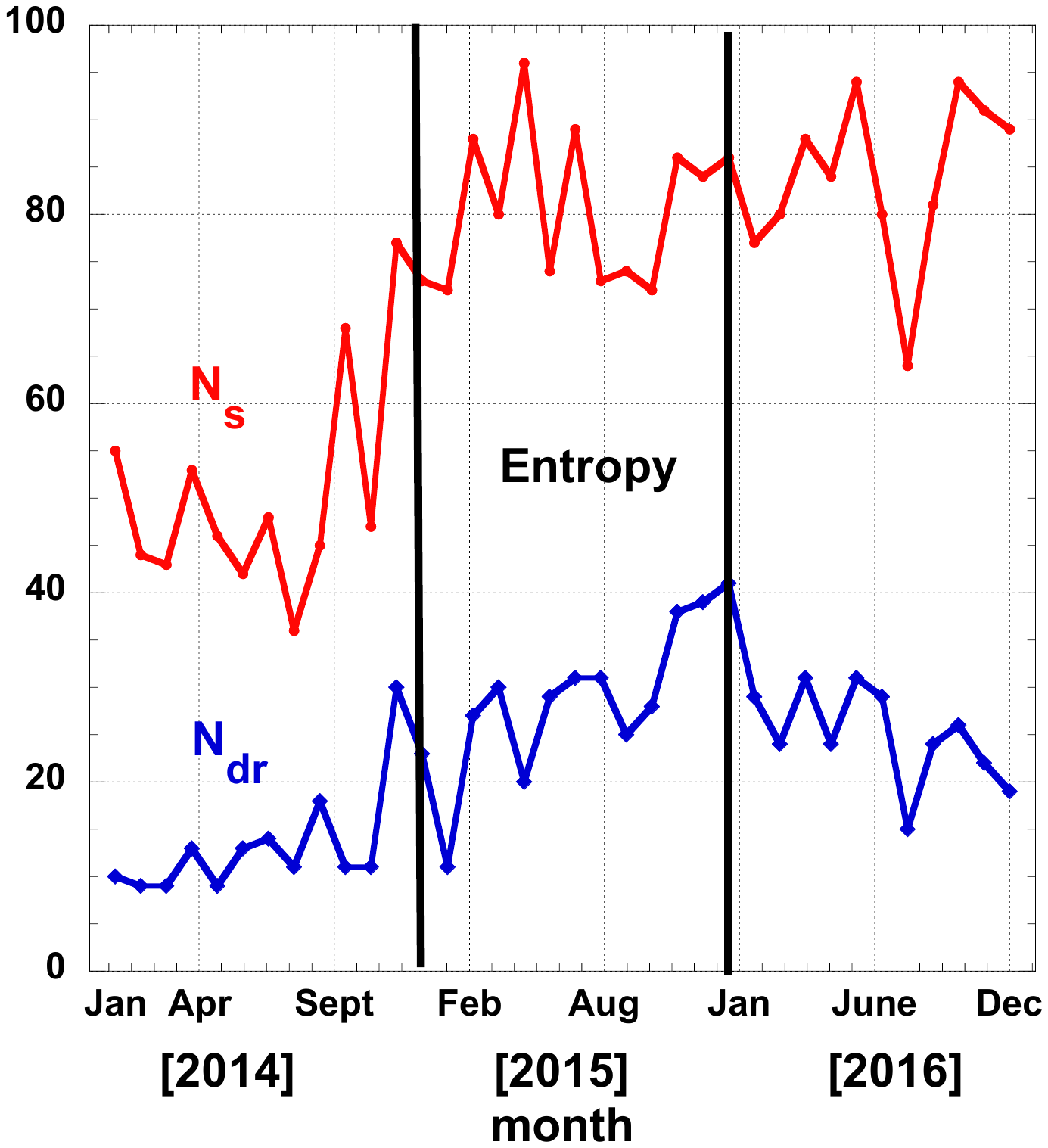}  
 \caption{  Entrance barrier load  conjecture effect. Visual correlation between  the number of  \underline{desk rejected} papers and the number of  submitted papers
 to  (left) {\it JSCS} during a given month,  between 2012 and  2014 or  to (right)  {\it Entropy}  during a given month between   2014  and  2016. } 
 \label{Plot5NdrNs} 
\end{figure}

A visual comparison  of the correlation between  the number of  \underline{desk rejected} papers and the number of  submitted papers
 to  {\it JSCS} during a given month, distinguishing  2013  from 2014 or  to   {\it Entropy}  during a given month in  2014 \underline{or} in 2015,  \underline{or} in 2016 is shown in Fig. \ref{Plot5NdrNs}.
For $JSCS$, the number of  desk rejected papers is roughly proportional to the number $N_s$ during a given month, $\simeq 25\%$, a value already noticed, - except at $N_{s}\sim 30$, when $N_{dr}$ can be as large as  30 - 50\%. However,  both in 2013 fall and 2014 spring-summer time, there are months for which $N_s$ is large, but $N_{dr}$  is low, leading to a doubt on a editor barrier load effect. 

For {\it Entropy},  it occurs that there are two clusters  separated by borders $N_s \sim 70$ and $N_{dr} \sim 20$. When  $N_s \ge 70$ , the number of desk rejected papers proportionally much increases. That was surely the case in 2015.

Conclusions: {\it JSCS} or $Entropy$  editors  may raise some entrance barrier   due to overload whatever the season,  

\section{  Optimal submission  month, - for paper later acceptance}\label{authorschance}

  The above data and discussion on the number of papers is relevant for editors, and automatic handling of papers.  Of course, this holds partially  true  as well for authors who do not want to overload editorial desks with many publications at a   given time, since authors expect some rather quick  (and positive) decision on their submission. However, another point is of great interest for authors, somewhat bearing on the reviewer and  desk editor mood. The most relevant question, on a possible seasonal bias, for authors is whether  his/her paper  has a greater chance to be accepted if submitted  during a given month. Thus, the  probability of acceptance, the so called "acceptance rate" is a relevant variable to be studied!

 The \underline{relative} number (i.e., monthly percentages) of papers accepted or rejected, $ p^{(m,y)}_{a,s}= N^{(m,y)}_a/N^{(m,y)}_s$ or $p^{(m,y)}_{r,s}=N^{(m,y)}_r/N^{(m,y)}_s$,   after submission on  a  specific  month  is   easily obtained from the figures.  
  The months ($mo$) can be ranked, e.g. in  decreasing order of importance, according to  such a relative probability (thereafter called $p_a$) of   having a   {paper accepted if submitted in a given month  ($m$)   to {\it JSCS} or to {\it Entropy} in  given years;  see  Table   \ref{JSCSEntropypamonth}. 
    One can easily obtained the corresponding $p_r$ of rejected papers; see Table \ref{JSCSEntropyprmonth}. This holds true  for any yearly time series leading to some  $p_a \equiv   p^{(m,y)}_{a,s}= N^{(m,y)}_a/N^{(m,y)}_s$,  whence allowing to compare journals according to
     \begin{equation}\label{eq1}
p_a-p_r =  \sum_y [p^{(m,y)}_{a,s}-p^{(m,y)}_{r,s} ]\equiv \sum_y  [\frac{N^{(m,y)}_a}{N^{(m,y)}_s}-\frac{N^{(m,y)}_r}{N^{(m,y)}_s}].
\end{equation} 

   One could also consider 
        \begin{equation}  \label{eq2}
q_a-q_r =    [\frac{\sum_y N^{(m,y)}_a}{\sum_y N^{(m,y)}_s}- \frac{\sum_y N^{(m,y)}_r}{\sum_y N^{(m,y)}_s} ] \equiv   [\frac{C^{(m)}_a}{C^{(m)}_s}- \frac{C^{(m)}_r}{C^{(m)}_s} ] 
\end{equation}
for the corresponding cumulated data over each specific time interval. A comment on the matter is postponed to the Appendix.

 \subsection{JSCS case}
  
   The  relevant  percentage  differences  between accepted and rejected number of papers  
   to {\it JSCS}  in  2013 and 2014 are given in   Fig. \ref{Plot23JSCSEntropydiffpapr12}.  
     
     From this difference in \underline{probability} perspective, it does not seem to be recommended that authors submit their paper to {\it JSCS}  in Mar or Dec.. They should rather submit their papers in 
     January, with some non-negligible statistical chance of acceptance for submissions  in February or October.

      \begin{figure} 
 \includegraphics[height=6.8cm,width=6.8cm]
{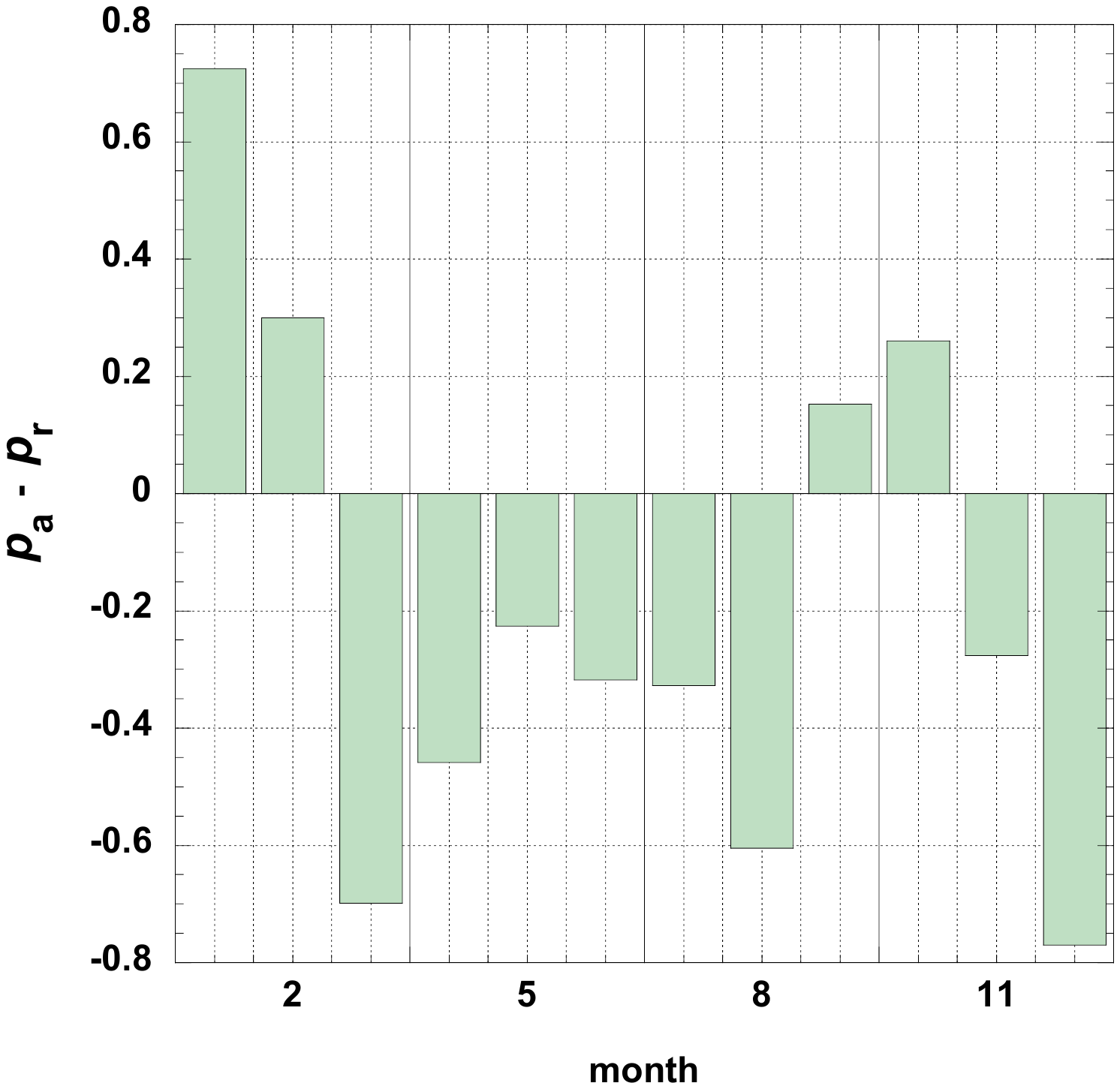} 
  \includegraphics [height=7.6cm,width=6.8cm]
{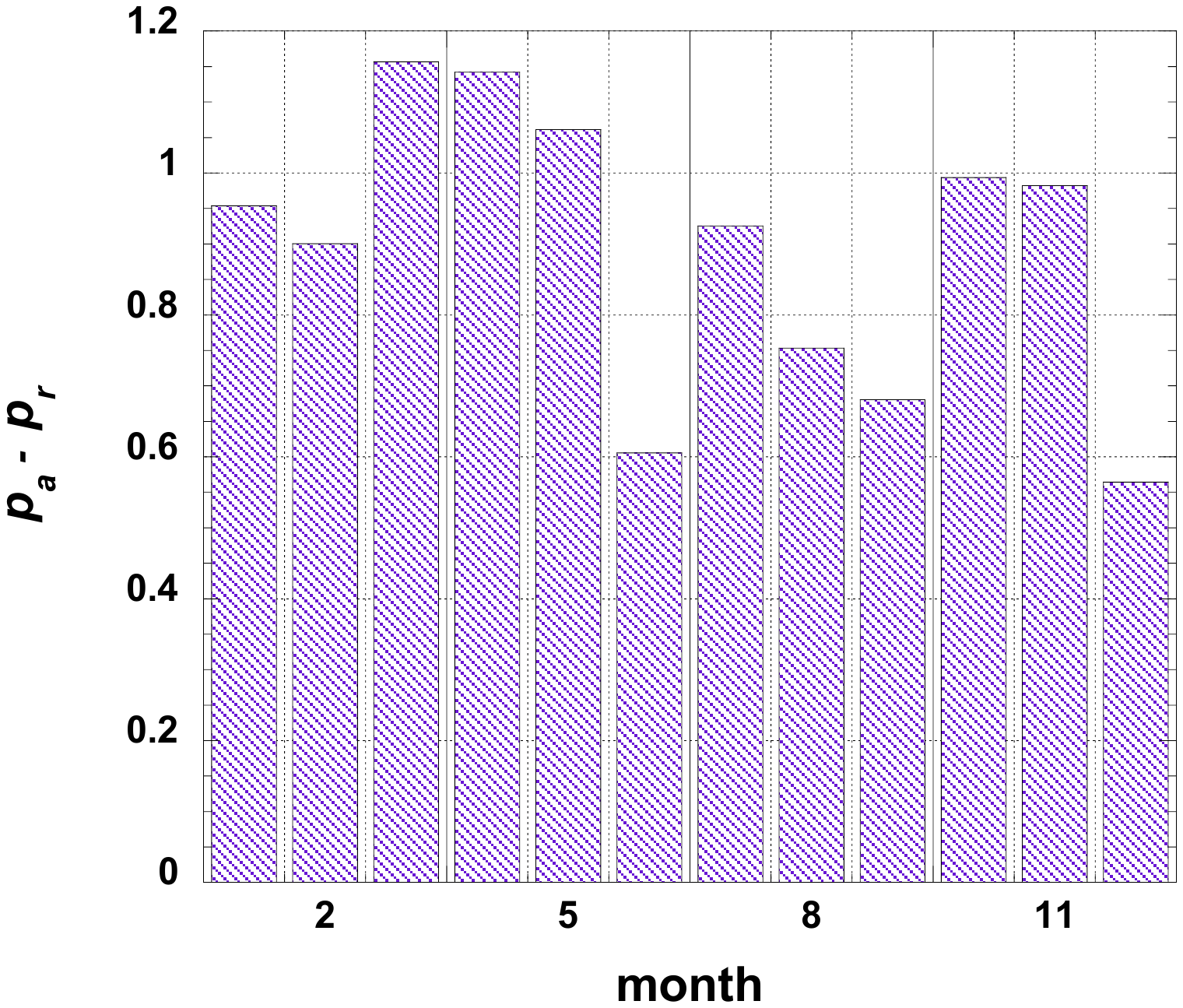}  
 \caption[ ] { Monthly aggregated percentages difference  between   accepted ($p_a$) and  rejected ($p_r$) number of papers, normalized 
 to the number of submitted papers,  on a given month  (left)  to {\it JSCS}  over [ 2012 - 2014], 
 (right)   to {\it Entropy}   in  [2014 -  2016].} 
   \label{Plot23JSCSEntropydiffpapr12} 
 \end{figure}  

 \subsection{Entropy case}
 For {\it Entropy},  an equivalent calculation of $p_a-p_r$ can be made, - from aggregating data in Fig. \ref{Plot8EntropyNsNatimeseries},  over  a 12 month interval leading to Fig. \ref{Plot23JSCSEntropydiffpapr12}.          Even though the  best percentage of accepted papers occurs if the papers are submitted from January till May (with a steady increase, in fact)  and in October and November, %
  the percentage of submitted papers in December is the largest of the year, and  the probability of acceptance is  the  lowest for such papers.  
      
       Thus, a marked  dip in acceptance probability occurs if the papers are submitted during the summer months [June-Sept.],  as seen in Fig. \ref{Plot23JSCSEntropydiffpapr12}, whence suggesting to avoid such months for submission to {\it Entropy}.

  \section{Warnings and Discussion} 
   For fully testing seasonal effects, one might argue that one should correlate the acceptance/rejection    matter to the hemisphere, or/and to nationality of authors, and notice the influence  of co-authors\footnote{Those of  editors might also be of concern:  most are Serbians for {\it JSCS}, the variety is large for {\it Entropy}. 
}.   

We apologize for not having searched for the  affiliations (either in the southern or northern hemisphere, - since seasons are different) of submitting authors to {\it Entropy};   we expect that  such a "hemisphere  effect", if it exists, is hidden in the statistical error bar of the sample, $ \simeq 1/\sqrt N_s \sim 4 \%$.  Concerning the nationalities of authors  (and reviewers)  of  {\it JSCS} in the period Nov. 2009 - Oct.  2014, those  have been discussed by  Nedic and Dekanski \cite{r9};  see Fig. 3 and Fig. 2 respectively in     \cite{r9}. For completeness, let us mention that the disitribution

  Data on  papers  submitted, accepted, rejected, withdrawn, to JSCS  from mainly Serbian authors and from "outside Serbia, on given years can be found in Table \ref{JSCSNsNaNw121314serbia}. From such a reading, it appears that JSCS editors are  fair, not biased, in favor or against papers with the corresponding author being from Serbia.

  At this level, 
more importantly, a comparison  resulting form the observation of   Fig. \ref{Plot23JSCSEntropydiffpapr12} allows to point to a marked difference between a specialized journal and a multidisciplinary one, - at least from the editorial aim, and the peer reviewers points of view.  The difference  between the  probability of acceptance and that of rejection, on a monthly basis,  i.e. $p_a-p_r$, 
has an astoundingly different behavior: the $p_a-p_r$ value is only positive over 3 months for {\it JSCS}, but is always positive for {\it Entropy}. This can be interpreted in terms of peer review control.  Recall that the percentage of desk rejection is approximatively the same for {\it JSCS} and {\it Entropy}, but the peer review rejection is much higher ($\sim 55\%$)  for {\it JSCS} in contrast with a $\sim 20\%$ reviewer rejection rate  for {\it Entropy}. In terms of seasonal effect, one has a positive value in January (and February) for {\it JSCS}, but  a positive effect  for the spring and fall months for {\it Entropy}. We consider that such a spread is likely due to the multidisciplinary nature of the  latter journal, reducing the strong monthly and seasonal bias on the fate of a paper.

 \section{Conclusion}\label{Conclusion}

Two remarks seem to be of interest for attempting some understanding of these  different findings. On one hand, statistical procedures  (either $\chi^2$ or confidence interval bounds $\mu \pm 2 \sigma$)  have not   to lead to identical conclusions: both can point to deviations, but  the former indicates  the presence (or absence) of peaks and dips with respect to the uniform distribution, while the latter  points to  statistical deviations when the distributions of residuals is expected to be like a Gaussian. In the latter case, an extension of the discussion including skewness and kurtosis is mandatory    
  \cite{r5}.  We have pointed out such departures from Gaussianity. The second remark on monthly  and/or seasonal bias, in view of the contradistinctions hereby found  between   the chemistry and multidisciplinary  journal,  might  not  be mainly due to desk rejection effects, as proposed by Shalvi et al.  \cite{r11}, but rather pertains to the peer reviewers  having different statuses  within the journal aims spread.

In so doing, by considering two somewhat "modest, but reliable" journals\footnote{
   The  value $T_a/T_s$ is often reported by publishers and editors as a "quality criterion" for a given journal. It is easy to find from the above data that  $T_a/T_s\simeq$ 0.44  and  0.49 for {\it JSCS} and {\it Entropy} respectively;  {\it JSCS} had  an impact factor  (IF) = 0.970 in  2015, which is the result of articles published in [2013-2014]; its 5-year IF   is 0.997, and h(2017) = 33;  {\it Entropy} had an IF = 1.821 (2016)  and a 5-year IF = 1.947 (2016), while h(2017) = 37.}
, we have demonstrated seasonal effects, in paper submission and also in subsequent acceptance. A seasonal bias effect is stronger in the specialized journal. Recall that one can usually read when an accepted paper has been submitted, but the missing set, the rejected papers when submitted, is usually unknown.  Due to our editorial status, we have provided a statistical analysis about such an information.  
Our  outlined findings and intrinsic behavioral hypotheses   
   markedly take into account the scientific work environment, and point, in the present cases, to seasonal  bias effects,  mainly due to authors in the submission/acceptance stage of the peer review process. 
   
  In order to go beyond our observation, we are aware that   more data must be made available by editors and/or publishers. Avoiding  debatable hypotheses on the quality of papers,  ranking of journals, fame of editors, open access publicity, submission fees, publication charges, and so on, we may suggest more work on time lag effects, beyond  Mrowinski et al.   \cite{r7,r8},    in order to pin point better the role of both editors and reviewers quality and concern.  
 In so doing, it might be wise to consider some ARCH-like modeling  of seasonal effects, as it has been done for observing day of the week effect in paper submission/acceptance/rejection to/in/by peer review journals  \cite{ausloos2017day}. This suggestion  of ARCH econometric-like modeling is supplementary supported by arguments as  in    related bibliometric studies. Indeed, one could develop a  Black--Scholes--Schr{\"o}dinger--Zipf--Mandelbrot model framework  for studying seasonal effects, - instead of  the coauthor core score  as in \cite{rotundo2014black}.

   \vskip0.5cm
\section*{Acknowledgements}
 MA greatly thanks the MDPI  {\it Entropy} Editorial staff for  gathering and cleaning up the raw data, and in particular
Yuejiao Hu, Managing Editor.
   \vskip0.5cm
   
{\bf \Large Appendix}
  \vskip0.5cm
In this Appendix, we discuss Eq.(\ref{eq2}), graphically displayed in Fig. \ref{Plot23JSCSEntropydiffqaqr12}. In some sense, this equation assumes that all years are equivalent and data for each month can be superposed whatever the year. It has been shown in the main text that such an aggregation process leads to a more comfortable statistical analysis.

  The $y$-axis scales appear to be markedly different in Fig. \ref{Plot23JSCSEntropydiffpapr12}  and  Fig. \ref{Plot23JSCSEntropydiffqaqr12}. However the patterns are very similar, thus {\it a priori} allowing for such an aggregation process. The conclusions on seasonal effects drawn from both figures or equations, Eq.(\ref{eq1}), and Eq.(\ref{eq2}), are therefore qualitatively similar.

      \begin{figure} 
 \includegraphics[height=6.8cm,width=6.8cm]
{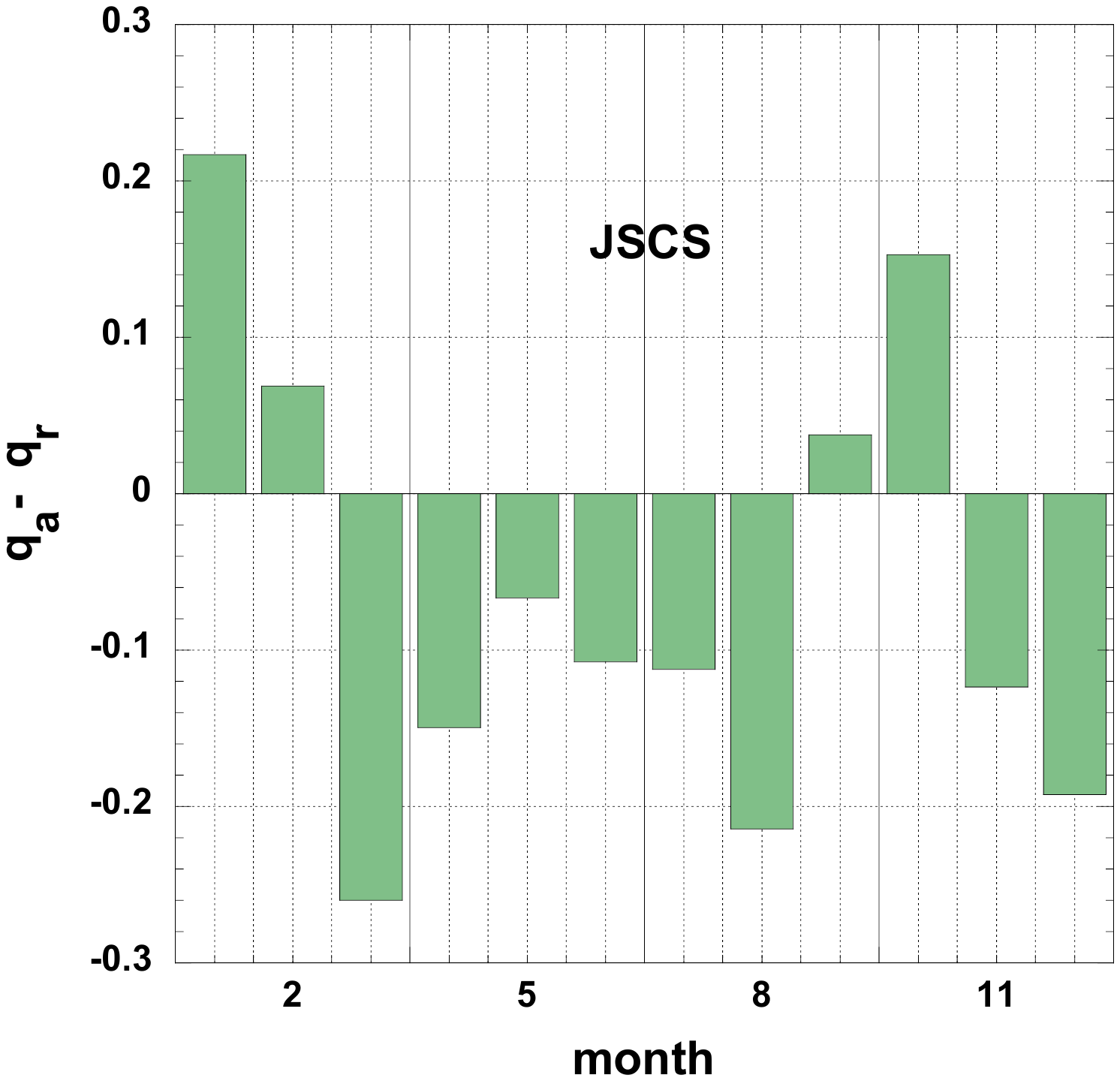} 
  \includegraphics [height=6.8cm,width=6.8cm]
{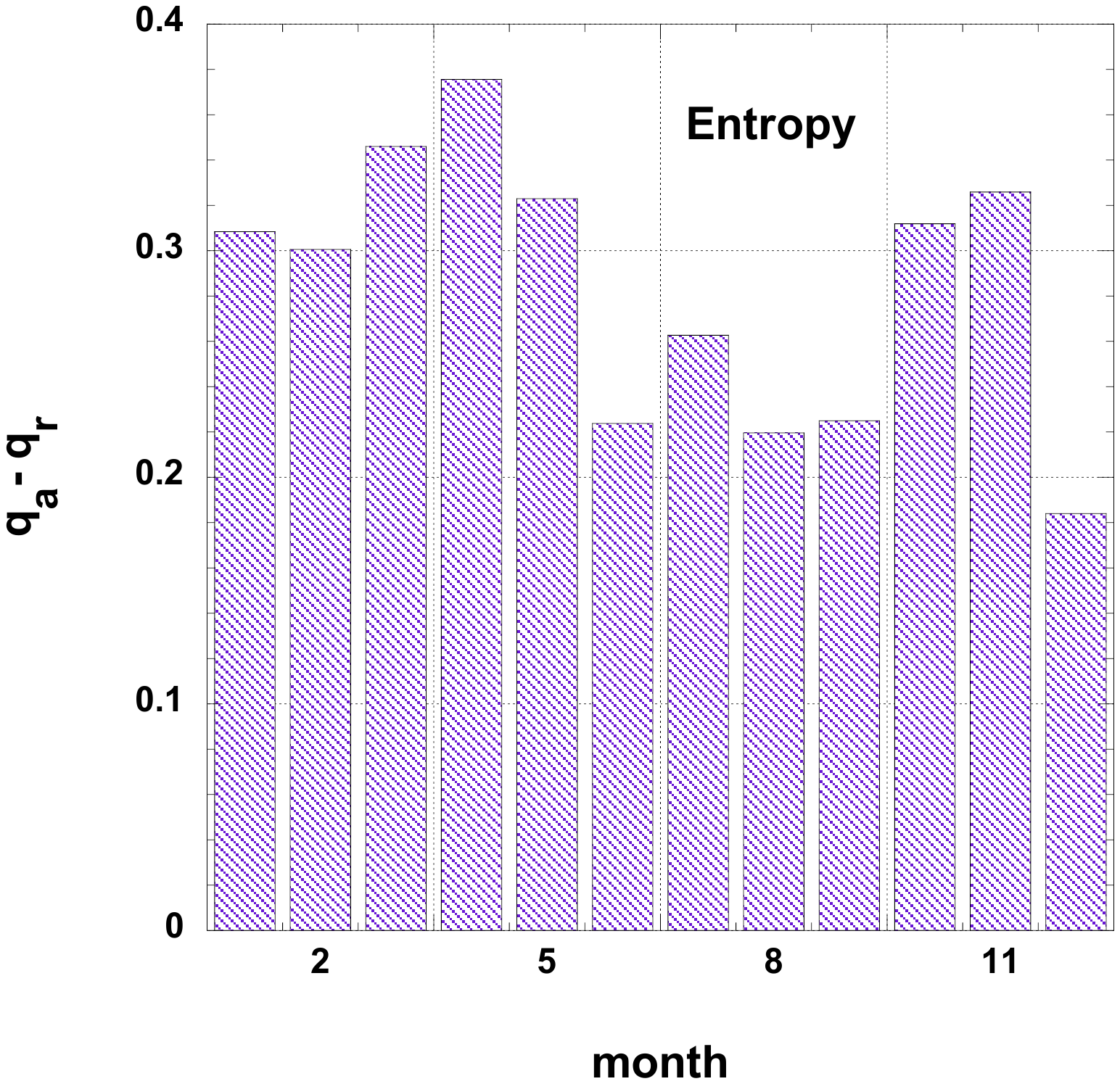}  
 \caption[ ] { Difference  between   accepted ($q_a\equiv [C_a^{(m)}/C_s^{(m])}]$) and  rejected ($q_r\equiv [C_r^{(m)}/C_s^{(m])}]$)  percentages of papers,  
 when the number of such papers is aggregated  in a given month for the examined time interval,    (left) over  [2012 -  2014]   to {\it JSCS}   
or   (right) over  [2014 -  2016]   to {\it Entropy.}}.
   \label{Plot23JSCSEntropydiffqaqr12} 
 \end{figure}

 \begin{table} 
 \caption{Statistical characteristics of the distribution of the Number $N_s^{(m,y)}$ of papers \underline{submitted in a given month}  to {\it JSCS} in  2012, 2013, in 2014, $C_s^{(m,y)}$,  in 2012,  2013 and 2014 and  $N_s^{(m,y)} $ over  [2013-2014];  notice that $C_s^{(m,y)}$ is obtained after monthly summing.  {\it  Therefore,  the statistical characteristics in the  last two columns slightly differ  from  each other,  because the time span  is determined  as occurring  over  $N.$ $mo=$ 12 or  36 months, respectively.}}  \begin{center}
 \label{JSCSNsmonthstat} 
      \begin{tabular}{|c|c|c|c|c|c|c|c|c|}
      \hline      
   {\it JSCS} &	$N_s^{(m,y)}$&$N_sr^{(m,y)}$&	$N_s^{(m,y)}$ & $C_s^{(m,y)}$&$N_s^{(m,y)} $ \\
 &$y= $ & $y= $&	$y= $&$y=(2012+$&   $y= [2012 $ \\
      & $  2012 $ & $2013$  &	$2014$ & $ 2013+2014)$&  $ -2014]$ 
        \\
     \hline      \hline 
Maximum	&	40	&	35	&	30	&	89 &	40	\\
Sum	&	317	&	322	&	274	&	913	&	913	\\
Points	&	12	&	12	&	12	&	12	&	36	\\
Mean	$\mu$&	26.417	&	26.833	&	22.833	&	76.083	&	25.361	\\
Median	&	26	&	28.50	&	24	&	79	&	26	\\
RMS	&	27.369	&	27.413	&	23.449	&	76.739	&	26.143	\\
Std Dev $\sigma$		&	7.4767	&	5.8595	&	5.5732	&	10.457	&	6.4372	\\
Variance	&	55.902	&	34.333	&	31.061	&	109.36	&	41.437	\\
Std Error	&	2.1583	&	1.6915	&	1.6088	&	3.0188	&	1.0729	\\
Skewness	&	0.46015	&	-0.4300	&	-0.1377	&	-0.5794	&	0.1897	\\
Kurtosis	&	-0.6336	&	-1.0609	&	-1.6261	&	-0.9949	&	-0.6144	 \\\hline
$\sigma/\mu$	&	0.2830	&	0.2184	&	0.2441	&	0.1374	&	0.2538	\\
$\mu-2\sigma$	&	11.464	&	15.114	&	11.687	&	55.169	&	12.757	\\
$\mu+2\sigma$	&	41.370	&	38.552	&	33.980	&	96.997	&	38.235	 \\\hline
$\chi^2 $&	23.278	&	14.075	&	14.964	&	15.811	&	57.186	\\
 $\chi^2_{N.mo-1}(0.95\%)$&\multicolumn{4}{|c|}{4.5748}&	22.465	\\  \hline
 \end{tabular}   \end{center}
  \end{table}   
 
 \begin{table}
 \caption{Statistical characteristics of the distribution of the Number $N_a^{(m,y)}$ of  \underline{accepted  papers}    \underline{if submitted in a given month ($m$)} to {\it JSCS} in  2012, 2013, in 2014, in  2012, 2013 and 2014 for $C_a^{(m,y)}$, and  over  [2013-2014];  $C_a^{(m,y)}$ is obtained after monthly summing. {\it  Therefore,  the statistical characteristics in the  last two columns slightly differ  from  each other,  because the time span  is determined  as occurring  over  $N.$ $mo=$ 12 or 36 months, respectively.}}    \begin{center}
  \label{JSCSNamonthstat}  
 \begin{tabular}{|c|c|c|c|c|c|c|c|c|c|}
      \hline      
   {\it JSCS} &	$N_a^{(m,y)}$&$N_a^{(m,y)}$&	$N_a^{(m,y)}$ &		$C_a^{(m,y)}$&$N_a^{(m,y)} $ \\
       &$y= $ & $y= $&	$y= $& 	  $y=(2012+$&   $y= [2012 $ \\
      & $  2012 $ & $2013$  &	$2014$ & $ 2013+2014)$&  $ -2014]$ \\
     \hline      \hline 
Min 	&	8 	&	8 	&	4 	&	27 	&	4 	\\
Max 	&	23 	&	20 	&	17	&	50 	&	23 	\\
Sum	&	160 	&	146 	&	116 	&	422 	&	422 	\\
Points	&	12 	&	12 	&	12 	&	12 	&	36 	\\
Mean$\mu$	&	13.333	&	12.167	&	9.6667	&	35.167	&	11.722	\\
Median	&	12.50 	&	10.50 	&	10 	&	33 	&	11 	\\
RMS	&	14.071	&	12.858	&	10.400	&	36.023	&	12.537	\\
Std Dev $\sigma$	&	4.6969	&	4.3450	&	4.0076	&	8.1557	&	4.5078	\\
Variance	&	22.061	&	18.879	&	16.061	&	66.515	&	20.321	\\
Std Error	&	1.3559	&	1.2543	&	1.1569	&	2.3543	&	0.7513	\\
Skewness	&	0.7156	&	0.6669	&	0.2111	&	0.7345	&	0.5718	\\
Kurtosis	&	-0.5251 	&	-1.1253	&	-0.9704	&	-0.7319	&	-0.3069	\\ \hline
\hline
$\sigma/\mu$& 0.3523 & 0.3571  &0.4146 & 0.2319 & 0.3846\\
$\mu-2\sigma$& 3.9396 &	3.4770	&	1.6520	&	18.855	&	2.7065 \\
$\mu+2\sigma$& 22.727 &	20.857	&	17.682	&	51.478	&	20.738  \\\hline
$\chi^2 $& 18.200	&17.068	&18.276	&20.806	&60.673 	\\ 
 $\chi^2_{N.mo-1}(0.95\%)$ &\multicolumn{4}{|c|}{4.5748}&	22.465	\\  \hline
 \end{tabular}    \end{center}
  \end{table}  
 
 \begin{table}  
 \caption{Statistical characteristics of the   distribution of the Number $N_r^{(m,y)}$ of  \underline{rejected papers}     \underline{if submitted in a given month ($m$)} to {\it JSCS} in 2012, 2013, and 2014,  after monthly summing,  for $C_r^{(m,y)}$, and  over  [2012-2014]. {\it  Therefore, the statistical characteristics in the  last two columns slightly differ  from  each other,  because the time span  is determined  as occurring  over  $N.$ $mo=$ 12 or 36 months, respectively.}}   \begin{center}
 \label{JSCSNrmonthstat} 
      \begin{tabular}{|c|c|c|c|c|c|c|c|c|c|}
      \hline      
   {\it JSCS} &	$N_r^{(m,y)}$&$N_r^{(m,y)}$&	$N_r^{(m,y)}$ &		$C_r^{(m,y)}$&$N_r^{(m,y)} $ \\
       & & &	 &	  $y=(2012+$&   $y= [2012 $ \\
      & $ y=2012 $ & $y=2013$  &	$y=2014$ & $ 2013+2014)$&  $ -2014]$ 
        \\
     \hline      \hline 
Min.&	6 	&	7	&	8 	&	27 	&	6 	\\
Max.&	21 	&	22 	&	24 	&	49 	&	24 	\\
Sum	&	153 	&	172 	&	153	&	478 	&	478 	\\
Points	&	12 	&	12 	&	12 	&	12 	&	36 	\\
Mean	$\mu$&	12.750	&	14.333	&	12.750 	&	39.833	&	13.278	\\
Median	&	13 	&	13.50 	&	12 	&	40 	&	13 	\\
RMS	&	13.488	&	15.067	&	13.432	&	40.507	&	14.016	\\
Std Dev $\sigma$	&	4.5950	&	4.8492	&	4.4133	&	7.6851	&	4.5520	\\
Variance	&	21.114	&	23.515	&	19.477	&	59.061	&	20.721	\\
Std Error	&	1.3264	&	1.3999	&	1.2740	&	2.2185	&	0.7587	\\
Skewness	&	0.2206	&	0.0324	&	1.3409	&	-0.2073	&	0.4918	\\
Kurtosis	&	-0.7684	&	-1.2912	&	1.4934	&	-1.3367	&	-0.5109	\\\hline				
\hline
$\sigma/\mu$&0.3604 &0.3383&0.3461& 0.1929 &0.3401\\
$\mu-2\sigma$&3.5601&	4.6348	&	3.9234	&	24.463	& 4.1738	\\
$\mu+2\sigma$&21.940&	24.032	&	21.577	&	55.204 	&	22.382   \\\hline
$\chi^2 $& 18.216 &18.047	& 16.804	&16.310	&54.619	\\ 
  $\chi^2_{N.mo-1}(0.95\%)$&\multicolumn{4}{|c|}{4.5748}&	22.465	\\  \hline
 \end{tabular}   \end{center}
 \end{table}   
 
  \begin{table} 
  \caption{  Statistical characteristics of the distribution of the Number $N_{dr}^{(m,y)}$ of  \underline{desk rejected papers}   
    \underline{if submitted in a given month ($m$)} to {\it JSCS} in 2012, 2013, in 2014, in 2012, 2013 and 2014 after monthly summing,  for $C_{dr}^{(m,y)}$, and  over  [2013-2014]. {\it  Therefore, the statistical characteristics in the  last two columns slightly differ  from  each other,  because the time span  is determined  as occurring  over  $N.$ $mo=$ 12 or  36 months, respectively.}} 
\begin{center}   \label{JSCSNdrmonthstat}
      \begin{tabular}{|c|c|c|c|c|c|c|c|}
      \hline      
   {\it JSCS} &	$N_{dr}^{(m,y)}$&$N_{dr}^{(m,y)}$&	$N_{dr}^{(m,y)}$ &		$C_{dr}^{(m,y)}$&$N_{dr}^{(m,y)} $ \\
       & $ y=$& $ y=$&$ y=$	 &	  $y=(2012+$&   $y= [2012 $ \\
      & $2012 $ & $2013$  &	$2014$ & $ 2013+2014)$&  $ -2014]$      \\
     \hline      \hline 
Min 	&	1 	&	2	&	4	&	9	&	1	\\
Max	&	8 	&	14	&	16	&	21	&	16	\\
Sum	&	42 	&	81	&	79	&	202	&	202	\\
Points	&	12 	&	12	&	12	&	12	&	36	\\
Mean	&	3.50	&	6.75	&	6.5833	&	16.833	&	5.6111	\\
Median	&	3 	&	7	&	5	&	18.50	&	5	\\
RMS	&	4.1433	&	7.6431	&	7.3428	&	17.378	&	6.5701	\\
Std Dev 	&	2.3160	&	3.7447	&	3.3967	&	4.5092	&	3.4664	\\
Variance	&	5.3636	&	14.023	&	11.538	&	20.333	&	12.016	\\
Std Error	&	0.6686	&	1.0810	&	0.9806	&	1.3017	&	0.5777\\
Skewness	&	0.5504	&	0.3994	&	1.9007	&	-0.6008	&	1.0552	\\
Kurtosis	&	-0.8601	&	-0.7819	&	3.0045	&	-1.1511	&	1.1309	\\ \hline			
$\sigma/\mu$& 0.6617& 0.5548 &0.5160 &0.2679 &0.6178\\
$\mu-2\sigma$&-1.1319& 	-0.739	&	-0.210	&	7.8148	&	-1.3217 \\
$\mu+2\sigma$&	8.1319& 14.239	&	13.377	&	25.852	&12.544  \\\hline
$\chi^2 $&	16.857	&22.852	&19.278	&13.287	&74.951\\	
 $\chi^2_{N.mo-1}(0.95\%)$&\multicolumn{4}{|c|}{4.5748}&	22.465	\\  \hline
 \end{tabular}   \end{center}	
  \end{table}

  \begin{table} 
  \caption{Statistical characteristics of the distribution of the Number $N_s^{(m,y)}$ of papers \underline{submitted in a given month ($m$)} to {\it Entropy} in year  ($y=$) 2014,  2015, and   2016,  after monthly summing for $C_s^{(m,y)}$, and  over  [2014-2016]. {\it  Therefore,  the statistical characteristics in the  last two columns slightly differ  from  each other,  because the time span  is determined  as occurring  over  $N.$ $mo=$ 12 or 36 months, respectively.}}   \begin{center} 
 \label{EntropyNsmonthstat} 
      \begin{tabular}{|c|c|c|c|c|c|c|c|c|}
      \hline      
  &$N_r^{(m,y)}$ &$N_r^{(m,y)}$  &	$N_r^{(m,y)}$	&$C_r^{(m,y)}$&  $N_r^{(m,y)} $  \\
    {\it Entropy}    &  & &	 &$y=(2014+$	 &\\
         & $y$=2014  &	$y$=2015&  $y$=2016&  $2015+ 2016)$& $y$=[2014-2016] \\
     \hline      \hline 
Min.	&	36 	&	72 	&	64 	&	173 	&	36 	\\
Max.	&	77 	&	96 	&	94 	&	250 	&	96 	\\
Total	&	604 	&	961 	&	1008 	&	2573 	&	2573 	\\
$N.$ $mo$	&	12&12&12&12&36\\
Mean ($\mu$)	&	50.33 	&	80.08 	&	84.00 	&	214.42	&	71.47 	\\
Median	&	46.5 	&	77.0 	&	85.0 	&	215.5 	&	75.5 	\\
RMS	&	51.548	&	80.473	&	84.388	&	215.26	&	73.608	\\
Std Dev($\sigma$) &	11.618	&	8.2623	&	8.4423	&	19.861	&	17.853	\\
Variance	&	134.97	&	68.265	&	71.273	&	394.45	&	318.71	\\
Std Err. 	&	3.3537	&	2.3851	&	2.4371	&	5.7333	&	2.9754	\\
Skewn.	&	1.2197	&	0.5534 	&	-0.9402	&	-0.3477 	&	-0.5598	\\
Kurt.&	0.5738 	&	-1.0378	&	0.6370 	&	0.1712	&	-1.0162	\\\hline
$\sigma/\mu$	&	0.2308 	&	0.1032	&	0.1005 	&	0.0926 	&	0.2498	\\
$\mu-2\sigma$	&	27.098	&	63.559	&	67.115	&	174.70	&	35.767	\\
$\mu+2\sigma$	&	73.569	&	96.608	&	100.88	&	254.14	&	107.18	\\\hline
$\chi^2 $ &29.4969	 &9.3767  &	9.3333 	&20.2356 	&156.075  \\
 $\chi^2_{N.mo-1}(0.95\%)$&\multicolumn{4}{|c|}{4.5748}&	22.465\\ \hline
 \end{tabular}   \end{center}
 \end{table}

   \begin{table} 
   \caption{Statistical characteristics of the distribution of the Number $N_a^{(m,y)}$ of  \underline{accepted  papers}    \underline{if submitted in a given month} to {\it Entropy} in 2014, in 2015,  and in 2016,  after monthly summing for $C_a^{(m,y)}$, and  over  [2014-2016]. {\it  Therefore,  the statistical characteristics in the  last two columns slightly differ  from  each other,  because the time span  is determined  as occurring  over  $N.$$m=$ 12 or 36 months, respectively.}} \begin{center}  
  \label{EntropyNamonthstat}
      \begin{tabular}{|c|c|c|c|c|c|c|c|c|}
      \hline      
  &$N_r^{(m,y)}$ &$N_r^{(m,y)}$  &	$N_r^{(m,y)}$	&$C_r^{(m,y)}$&  $N_r^{(m,y)} $  \\
    {\it Entropy}    &  & &	 &$y=(2014+$	 &\\
         & $y$=2014  &	$y$=2015&  $y$=2016&  $2015+ 2016)$& $y$=[2014-2016] \\
     \hline      \hline 
Minimum	&	16 	&	27 	&	30 	&	77 	&	16 	\\
Maximum	&	44 	&	56 	&	51 	&	123 	&	56 	\\
Total	&	336 	&	467 	&	447 	&	1250 	&	1250	\\
$N.$$m$	&	12 	&	12 	&	12 	&	12 	&	36 	\\
Mean ($\mu$)		&	28.00 	&	38.92	&	37.25 	&	104.17	&	34.72 	\\
Median	&	30.5 	&	38.5  	&	36.0 	&	103.5 	&	34.5 	\\
RMS	&	28.960	&	39.530	&	37.737	&	104.98	&	35.709	\\
Std Dev($\sigma$)	&	7.7225	&	7.2420	&	6.3120	&	13.617	&	8.4536	\\
Variance	&	59.636	&	52.447	&	39.841	&	185.42	&	71.463	\\
Std Err 	&	2.2293	&	2.0906	&	1.8221	&	3.9309	&	1.4089	\\
Skewn.	&	0.21896	&	0.82477	&	0.76955	&	-0.43449	&	0.05694 	\\
Kurt.	&	-0.16185	&	0.96295	&	-0.18015	&	-0.54925	&	0.38648	\\\hline
$\sigma/\mu$	&	0.2758	&	0.1861	&	0.1694 	&	0.1307 	&	0.2435	\\
$\mu-2\sigma$	&	12.555	&	24.433	&	24.626	&	76.933	&	17.815	\\
$\mu+2\sigma$	&	43.445	&	53.401	&	49.874	&	131.40	&	51.629	\\\hline
$\chi^2$	&	23.4286	&14.8243	&11.7651 	&19.5802	&72.0357	 \\
 $\chi^2_{N.mo-1}(0.95\%)$&\multicolumn{4}{|c|}{4.5748}&	22.465\\ \hline
 \end{tabular}   \end{center}
  \end{table}  
 
  \begin{table} \begin{center}
  \caption{Statistical characteristics of the distribution of the Number $N_r^{(m,y)}$  of  \underline{rejected papers}     \underline{if submitted in a given month} to {\it Entropy}  in 2014, in 2015, and in 2016, after monthly summing for $C_r^{(m,y)}$, and  over  [2014-2016]. {\it  Therefore, the statistical characteristics in the  last two columns slightly differ  from  each other,  because the time span  is determined  as occurring  over  $N.$ $mo=$ 12 or 36 months, respectively.}}  
 \label{EntropyNrmonthstat}
      \begin{tabular}{|c|c|c|c|c|c|c|c|c|}
      \hline      
  &$N_r^{(m,y)}$ &$N_r^{(m,y)}$  &	$N_r^{(m,y)}$	&$C_r^{(m,y)}$&  $N_r^{(m,y)} $  \\
    {\it Entropy}    &  & &	 &$y=(2014+$	 &\\
         & $y$=2014  &	$y$=2015&  $y$=2016&  $2015+ 2016)$& $y$=[2014-2016] \\
     \hline      \hline 
Min.	&	3 	&	8 	&	13 	&	32 	&	3 	\\
Max.	&	14 	&	21 	&	32 	&	58 	&	32 	\\
Total	&	110 	&	162 	&	246 	&	518 	&	518 	\\
$N.$ $mo$	&	12 	&	12 	&	12 	&	12 	&	36 	\\
Mean ($\mu$)	&	9.1667	&	13.500	&	20.500	&	43.167	&	14.389	\\
Median	&	9.5 	&	12.0 	&	19.5 	&	43.0 	&	13.0 	\\
RMS	&	9.8826	&	14.160	&	21.264	&	43.648	&	15.815	\\
Std Dev($\sigma$)	&	3.8573	&	4.4620	&	5.9007	&	6.7532	&	6.6559	\\
Variance	&	14.879	&	19.909	&	34.818	&	45.606	&	44.302	\\
Std Err 	&	1.1135	&	1.2881	&	1.7034	&	1.9495	&	1.1093	\\
Skewn.	&	-0.3109	&	0.5387 	&	0.3993 	&	0.5855	&	0.5970	\\
Kurt.	&	-1.3129	&	-1.1113	&	-0.6683 	&	0.2828 	&	0.0303  	\\\hline
$\sigma/\mu$	&	0.4208 	&	0.3305 	&	0.2878 	&	0.1564 	&	0.4626	\\
$\mu-2\sigma$	&	1.4521	&	4.5761	&	8.6986	&	29.660	&	1.0770	\\
$\mu+2\sigma$	&	16.881	&	22.424	&	32.301	&	56.673	&	27.701	\\\hline
$\chi^2 $ &	17.8545	&16.2222 	&18.6829 	&11.6215 	&107.7598 \\
 $\chi^2_{N.mo-1}(0.95\%)$&\multicolumn{4}{|c|}{4.5748}&	22.465\\ \hline
 \end{tabular}   \end{center}
  \end{table}

 \begin{table} \begin{center}
 \caption{Statistical characteristics of the distribution of the Number $N_{dr}^{(m,y)}$  of  \underline{ desk rejected  papers}    \underline{if submitted in a given month ($m$)} to {\it Entropy}  in 2014, in 2015, and in 2016, after monthly summing for $C_{dr}^{(m,y)}$, and  over  [2014-2016]. {\it  Therefore,  the statistical characteristics in the  last two columns slightly differ  from  each other,  because the time span  is determined  as occurring  over  $N.$ $mo=$ 12 or 36 months, respectively.}}  
  \label{EntropyNdrmonthstat}
      \begin{tabular}{|c|c|c|c|c|c|c|c|c|}
      \hline      
  &$N_{dr}^{(m,y)}$ &$N_{dr}^{(m,y)}$  &	$N_{dr}^{(m,y)}$	&$C_{dr}^{(m,y)}$&  $N_{dr}^{(m,y)} $  \\
    {\it Entropy}    &  & &	 &$y=(2014+$	 &\\
         & $y$=2014  &	$y$=2015&  $y$=2016&  $2015+2016)$& $y$=[2014-2016] \\
     \hline      \hline 
Min.	&	9	&	11 	&	15 	&	49 	&	9 	\\
Max.	&	30 	&	39 	&	41 	&	88 	&	41 	\\
Total	&	158 	&	332 	&	315 	&	805 	&	805 	\\
$N.$ $mo$	&	12 	&	12 	&	12 	&	12 	&	36 	\\
Mean ($\mu$)	&	13.167	&	27.667	&	26.250	&	67.083	&	22.361	\\
Median	&	11.0 	&	28.5 	&	25.0	&	69.0 	&	24.0 	\\
RMS	&	14.329	&	28.601	&	27.017	&	67.894	&	24.175	\\
Std Dev. ($\sigma$)	&	5.9058	&	7.5719	&	6.6759	&	10.925	&	9.3171	\\
Variance	&	34.879	&	57.333	&	44.568	&	119.36	&	86.809	\\
Std Err 	&	1.7049	&	2.1858	&	1.9272	&	3.1538	&	1.5529	\\
Skewness	&	2.1223	&	-0.5584	&	0.4944	&	0.0229  	&	0.0697  	\\
Kurtosis	&	3.6775	&	0.2982 	&	0.3776 	&	-0.5016 	&	-1.0381	\\\hline
$\sigma/\mu$	&	0.4485 	&	0.2737	&	0.2543 	&	0.1629	&	0.4167	\\
$\mu-2\sigma$	&	1.3550	&	12.523	&	12.898	&	45.233	&	3.7269	\\
$\mu+2\sigma$	&	24.978	&	42.810	&	39.602	&	88.933	&	40.995	\\\hline
$\chi^2 $ &21.5197   & 10.0404 	& 8.2881  & 	6.5221	& 15.5365  \\
 $\chi^2_{N.mo-1}(0.95\%)$&\multicolumn{4}{|c|}{4.5748}&	22.465\\ \hline
 \end{tabular}   \end{center}
  \end{table}

  \begin{table} 
  \caption{Months ($mo$) ranked in  decreasing order of importance according to the probability $p_a  =  N_a^{(m)} /N_s^{(m)}$  of   having a   \underline{paper accepted}   \underline{if submitted in a given month} ($m$)   to {\it JSCS} or to {\it Entropy} in  given years. }    \label{JSCSEntropypamonth}
   \begin{center}
   \begin{tabular}{|c|c|c|c|c|c| |c|c|c|c|c|c|}
      \hline    
         \multicolumn{6}{|c||}{{\it JSCS}} &      \multicolumn{6}{|c|}{{\it Entropy}} 	\\ \hline       
      \multicolumn{2}{|c|}{2012}&  \multicolumn{2}{|c|}{2013}&	\multicolumn{2}{|c||}{2014} & \multicolumn{2}{|c|}{2014} & \multicolumn{2}{|c|}{2015}& \multicolumn{2}{|c|}{2016}	\\ \hline
$p_a$& $mo$  &	$p_a $&$mo$& $p_a $&$mo$& $p_a $&$mo$ & $p_a $&$mo$ & $p_a $&$mo$ \\  
  \hline      \hline
0.6923	&	Jan	&	0.6400	&	Nov 	&	0.6364	&	Jan 	&	0.6977	&	Mar	&	0.5833	&	Feb 	&	0.5604	&	Nov 	\\
0.6191	&	Sep	&	0.6250	&	Oct 	&	0.5862	&	Apr	&	0.6739	&	May	&	0.5833	&	May	&	0.5000	&	Apr 	\\
0.6053	&	Oct	&	0.5882	&	Feb 	&	0.4815	&	July	&	0.6471	&	Oct 	&	0.5479	&	Jan 	&	0.4687	&	Aug 	\\
0.6000	&	Feb	&	0.5312	&	Sept 	&	0.4737	&	June	&	0.6458	&	July	&	0.5405	&	Sept 	&	0.4545	&	Feb 	\\
0.5909	&	Aug	&	0.5143	&	Jan 	&	0.4615	&	Feb 	&	0.5849	&	Apr	&	0.5227	&	Mar 	&	0.4500	&	Mar 	\\
0.5385	&	Nov	&	0.5000	&	May	&	0.4444	&	Aug 	&	0.5818	&	Jan 	&	0.5000	&	Oct 	&	0.4468	&	June	\\
0.4839	&	July	&	0.4167	&	June	&	0.4074	&	Sept 	&	0.5556	&	Aug 	&	0.5000	&	Apr	&	0.4362	&	Oct 	\\
0.4750	&	Dec	&	0.3810	&	Dec	&	0.4000	&	Oct 	&	0.5532	&	Nov 	&	0.4595	&	June	&	0.4286	&	May	\\
0.4615	&	May	&	0.3667	&	Aug 	&	0.3929	&	Mar	&	0.5455	&	Feb 	&	0.4186	&	Nov 	&	0.4270	&	Dec	\\
0.4091	&	June	&	0.3548	&	July	&	0.3750	&	May	&	0.4286	&	Dec	&	0.4157	&	July	&	0.4000	&	July	\\
0.3548	&	Apr	&	0.2963	&	Apr 	&	0.2353	&	Dec	&	0.4000	&	Sept 	&	0.3929	&	Dec	&	0.3827	&	Sept 	\\
0.3158	&	Mar	&	0.2667	&	Mar 	&	0.1667	&	Nov 	&	0.3810	&	June	&	0.3699	&	Aug 	&	0.3605	&	Jan 	\\
\hline \end{tabular}   \end{center} 
 \end{table}

  \begin{table} 
  \caption{Months ($mo$) ranked in  decreasing order of importance according to the probability $p_r  =  N_r^{(m)} /N_s^{(m)}$  of   having a   \underline{paper rejected}   \underline{if submitted in a given month} ($m$)   to {\it JSCS} or to {\it Entropy} in  given years. }    \label{JSCSEntropyprmonth}  \begin{center} 
   \begin{tabular}{|c|c|c|c|c|c| |c|c|c|c|c|c|}
      \hline    
         \multicolumn{6}{|c||}{{\it JSCS}} &      \multicolumn{6}{|c|}{{\it Entropy}} 	\\ \hline       
      \multicolumn{2}{|c|}{2012}&  \multicolumn{2}{|c|}{2013}&	\multicolumn{2}{|c||}{2014} & \multicolumn{2}{|c|}{2014} & \multicolumn{2}{|c|}{2015}& \multicolumn{2}{|c|}{2016}	\\ \hline
$p_r$& $mo$  & $p_r$& $mo$  &	$p_r $&$mo$& $p_r $&$mo$& $p_r $&$mo$ & $p_r $&$mo$ \\  
   \hline      \hline																			
0.6129	&	Apr		&	0.7333	&	Mar		&	0.8000	&	Nov		&	0.3095	&	June		&	0.2639	&	 Feb 	&	0.3596	&	Dec \\
0.5909	&	June		&	0.7037	&	Apr		&	0.7647	&	Dec		&	0.2500	&	  Feb  	&	0.2360	&	July		&	0.3210	&	 Sept  \\
0.5263	&	Mar		&	0.6333	&	 Aug 	&	0.6000	&	 Oct		&	0.2364	&	 Jan 	&	0.2083	&	May		&	0.2969	&	 Aug  \\
0.5250	&	Dec 		&	0.6129	&	July		&	0.5714	&	Mar		&	0.2128	&	Nov		&	0.2055	&	 Aug 	&	0.2872	&	 Oct	\\
0.5161	&	July		&	0.5714	&	Dec		&	0.5625	&	May		&	0.2000	&	 Sept 	&	0.1705	&	Mar		&	0.2857	&	May	\\
0.5000	&	May		&	0.5000	&	June		&	0.5556	&	 Aug 	&	0.1912	&	 Oct		&	0.1486	&	June		&	0.2500	&	Mar	\\
0.4615	&	Nov		&	0.5000	&	May		&	 0.5556	&	 Sept 	&	0.1818	&	Dec		&	0.1429	&	Dec		&	0.2375	&	July	\\
0.4000	&	 Feb 	&	0.4857	&	 Jan 	&	0.5385	&	 Feb 	&	0.1698	&	Apr		&	0.1395	&	Nov		&	0.2234	&	June	\\
0.3947	&	 Oct		&	0.4687	&	 Sept 	&	0.5263	&	June		&	0.1389	&	 Aug 	&	0.1370	&	 Jan 	&	0.1978	&	Nov \\
0.3809	&	 Sept 	&	0.4118	&	 Feb 	&	0.5185	&	July		&	0.1304	&	May		&	0.1250	&	Apr		&	0.1688	&	 Feb  \\
0.3636	&	 Aug 	&	0.3750	&	 Oct		&	0.3793	&	Apr		&	0.0930	&	Mar		&	0.1216	&	 Sept 	&	0.1628	&	 Jan  \\
0.3077	&	 Jan 	&	0.3600	&	Nov		&	0.3636	&	 Jan 	&	0.0625	&	July		&	0.1111	&	 Oct		&	0.1477	&	Apr\\ \hline
 \end{tabular}   \end{center} 
 \end{table} 
 
   \begin{table}
 \caption{Data on  papers  submitted, accepted, rejected, withdrawn, to JSCS  from mainly Serbian authors and from "outside Serbia", on given years.  }    \begin{center}
  \label{JSCSNsNaNw121314serbia}  
 \begin{tabular}{|c|c|c|c|c|c|c|c|c|c|}
 \hline
 $JSCS$& papers  & $Total$& $from$  &	$from$  \\  
   &  &  & $Serbia$  &	$"outside" $  \\  
     \hline       2012	& Submitted&	317	&	84	&	233	\\	
&	Accepted	&	160	&	70	&	90	\\	
&	Rejected	&	153	&	14	&	139	\\	
&	Withdrawn	&	4	&	1	&	3	\\	
				\hline					
2013	&Submitted&	322	&	92	&	231	\\	
&	Accepted	&	146	&	65	&	81	\\	
&	Rejected	&	172	&	25	&	147	\\	
&	Withdrawn	&	4	&	2	&	2	\\	
				\hline					
2014	&Submitted&	293	&	83	&	210	\\	
&	Accepted	&	130	&	62	&	68	\\	
&	Rejected	&	160	&	20	&	140	\\	
&	Withdrawn	&	3	&	1	&	2	\\	
  \hline
 \end{tabular}    \end{center}
  \end{table}  


\clearpage

\begin{thebibliography}{9}


\bibitem{Nedicefficiency}
Nedić, O., Drvenica, I., Ausloos, M., \& Dekanski, A. B. (2018). Efficiency in managing peer-review of scientific manuscripts–editors’ perspective. {\it J. Serb. Chem. Soc.} {\bf 83}, 1391-1405.


\bibitem{Nedicauthorsperspective}
Drvenica, I., Bravo, G., Vejmelka, L., Dekanski, A., \& Nedić, O. (2019). Peer Review of Reviewers: The Author’s Perspective.  {\it Publications} {\bf 7},  1.


\bibitem{SCIMHerteliu}
Boja, C. E., Herţeliu, C., Dârdală, M., \& Ileanu, B. V. (2018). Day of the week submission effect for accepted papers in Physica A, PLOS ONE, Nature and Cell. {\it Scientometrics} {\bf 117}, 887-918.


\bibitem{r11}
  Shalvi, S., Baas, M., Handgraaf,  M.J.J,, \& De Dreu, C.K.W.  (2010).  "Write when hot - submit when not: seasonal bias in peer review or acceptance?" {\it Learned Publishing}  {\bf 23},  117-123. 

\bibitem{r10}
 Schreiber, M.  (2012).  "Seasonal bias in editorial decisions for a physics journal: you should write when you like, but submit in July".  {\it Learned Publishing}  {\bf  25}, 145-151.

\bibitem{r1}
 Alikhan, A., Karan,  A.,  Feldman, S.R., \& Maibach, H.I.   (2011).
   "Seasonal variations in dermatology manuscript submission".  {\it Journal of Dermatological Treatment}
\textbf{22},  60.  

\bibitem{r2}
 Ausloos,  M., Nedic, O., \&    Dekanski,  A.  (2016).
 "Day of the week effect in paper submission/acceptance/rejection to/in/by peer review journals".  \textit{Physica A} \textbf{456}, 197-203.

\bibitem{r9}
 Nedic, O. \&    Dekanski, A. (2015). "A survey on publishing policies of the Journal of the Serbian Chemical Society -  On the occasion of the 80th volume". {\it J. Serb. Chem. Soc.}  {\bf  80},  959-969. 

\bibitem{r3}
  Callaham, M.L., Baxt, W.J., Waeckerie,  J.F.,  \& Wears, R.L.  (1998). "Reliability of Editor's subjective quality ratings of Peer reviews of manuscripts".  {\it JAMA}, {\bf  280}, 229-231,  and refs therein.   
  
\bibitem{r4}
  Cole, S., Cole, J.R.,  \&  Simon, G.A.   (1981).   "Chance and Consensus in Peer Review". {\it Science} {\bf  214}, 881-886.
 
\bibitem{r6}
 Hargens, L.L.  (1968).
  "Scholarly Consensus and Journal Rejection Rates". {\it American Sociological Review}  {\bf  53},  139-151.
  
\bibitem{r5}
 Doane, D.P.  \& Seward, L.E.   (2011),   {\it  Applied Statistics in Business and Economics, 3rd ed}., McGraw-Hill/Irwin pp. 154-156.
 
 \bibitem{r7}  Mrowinski, M.J.,   Fronczak, A.,  Fronczak, P.,    Nedic, O.,   \&  Ausloos, M.    (2016). "Review times in peer review: quantitative analysis and modelling of editorial work flows", {\it Scientometrics}  {\bf  107}, 271-286.

 \bibitem{r8} Mrowinski, M.J.,   Fronczak, A.,  Fronczak, P.,    Nedic, O.,    \&  Ausloos, M.    (2017).   "Artificial intelligence in peer review: how can evolutionary computation support journal editors?". {\it PLOS ONE} \textbf{12},  e0184711.
  
     \bibitem{ausloos2017day}  Ausloos,  M,, Nedic, O.,   Dekanski, A., Mrowinski, M.J.,   Fronczak, A.,  \& Fronczak, P, (2017).  "Day of the week effect in paper submission/acceptance/rejection to/in/by peer review journals. II. An ARCH econometric-like modeling",
  {\it Physica A: Statistical Mechanics and its Applications}, {\bf 468}, 462--474.
 
   \bibitem{rotundo2014black} Rotundo, G. (2014).  "Black--Scholes--Schr{\"o}dinger--Zipf--Mandelbrot model framework for improving a study of the coauthor core score".  {\it Physica A: Statistical Mechanics and its Applications},
{\bf 404}, 296--301.
   
\end{thebibliography}
 \end{document}